\documentclass{article}
\usepackage[left=1.5in, right=1.5in, bottom=1in, top=1in]{geometry}
\usepackage{graphicx}
\usepackage{authblk}
\usepackage{soul}
\usepackage{amsmath,commath,color}
\usepackage{multirow}
\usepackage{commath}
\usepackage{textcomp}
\usepackage{gensymb}
\usepackage{amssymb}

\newcommand{\bs}{\boldsymbol}

\title{Oriented suspension mechanics with applications to flow linear dichrosim spectroscopy and pathogen detection}

\author[1]{G. Cupples}
\author[1]{D. J. Smith}
\author[2]{M. R. Hicks}
\author[1,*]{R. J. Dyson}

\affil[1]{\small School of Mathematics, University of Birmingham, B15 2TT, U.K.}
\affil[2]{\small Linear Diagnostics Ltd, BioHub Birmingham, 97 Vincent Drive, Birmingham, B15 2SQ, U.K.}
\affil[*]{\small Correspondence email: r.j.dyson@bham.ac.uk}

\begin{document}
\maketitle

\begin{abstract}
Flow linear dichroism is a biophysical spectroscopic technique that exploits the shear-induced alignment of elongated particles in suspension. Motivated by the broad aim of optimising the sensitivity of this technique, and more specifically by a handheld synthetic biotechnology prototype for waterborne-pathogen detection, a model of steady and oscillating pressure-driven channel flow and orientation dynamics of a suspension of slender microscopic fibres is developed. The model couples the Fokker-Planck equation for Brownian suspensions with the narrow channel flow equations, the latter modified to incorporate mechanical anisotropy induced by the particles. The linear dichroism signal is estimated through integrating the perpendicular components of the distribution function via an appropriate formula which takes the bi-axial nature of the orientation into account. For the specific application of pathogen detection via binding of M13 bacteriophage, it is found that increases in the channel depth are more significant in improving the linear dichroism signal than increases in the channel width. Increasing channel depth to \(2\)~mm and pressure gradient to $5\times10^4~$Pa/m essentially maximises the alignment. Oscillating flow can produce nearly equal alignment to steady flow at appropriate frequencies, which has significant potential practical value in the analysis of small sample volumes. 
\end{abstract}

\section{Introduction}


Suspensions of particles in liquid or gas are found throughout the natural world and in many industrial processes; including blood, particulate-laden air, algae in open water or in bioconvection experiments, semen samples, and industrial applications such as emulsions in food and cosmetics -- these examples and more are elaborated in references \cite{kim2013microhydrodynamics,hill2005bioconvection,creppy2016,guazzelli2011physical}. We will consider flowing suspensions of microscopic elongated rod-like particles, as a means to understand and improve flow linear dichroism spectroscopy, with particular focus on a specific technological application. The core idea of flow linear dichroism spectroscopy is that elongated particles undergo shear-induced rotation, which has the effect of concentrating their orientation in the direction of flow. The difference in absorbance of light polarised parallel, and perpendicular to, the flow direction can be used to reveal structural information; further information and wider context is given in references \cite{marrington2005validation,daviter2013circular,bulheller2007circular,norden1997circular}. The application of mathematical modelling to understand and quantify molecular orientation in flow linear dichroism was addressed by McLachlan et al.\ \cite{mclachlan2013calculations}, building on classical oriented suspension mechanics \cite{batchelor1971stress,strand1987computation}. This manuscript will generalise this work to the significantly more complex problem of a non-homogeneous shear environment of pressure-driven, and potentially time-varying, channel flow, and will moreover focus on a specific technological application.

The specific application is a prototype handheld device, developed by Linear Diagnostics Ltd (LD) designed to detect waterborne pathogens in fluids. The analyte is mixed with a reagent containing a synthetic biology micron-length fibre based on M13 bacteriophage. In the absence of the pathogen, the fibres align in the flow field and therefore exhibit linear dichroism. The M13 are engineered with antibodies enabling binding to specific pathogens, therefore in the presence of the pathogen the alignment is disrupted, resulting in a reduced linear dichroism signal. The technique is described further in reference \cite{pacheco2011detection}. To improve the sensitivity of the device it is desirable to optimise the signal to noise ratio, which can be achieved by increasing the overall alignment in the system. To analyse the system theoretically involves solving a coupled pressure-driven steady or oscillating flow and alignment problem. While we present results for a specific biotechnology application, the computational framework is generic to any channel flow of dilute suspensions.

The coupling between the flow and particle orientation is important to understand the dynamics of these suspensions. Jeffrey \cite{jeffery1922motion} carried out a landmark study of the dynamics of suspended elongated particles in viscous flow, in particular shear induced alignment. Moreover, following seminal theoretical work \cite{ericksen1960ti,leslie1968some,batchelor1970stress,peterlin1939}, it has been established how the aspect ratio of particles affects emergent rheology. In the related context of suspensions of actively-swimming oriented particles, Pedley \& Kessler \cite{pedley1990new} described a coupled model of orientation distribution, governed by a Fokker-Planck equation, and fluid flow, given by modifications to the Navier-Stokes equations proportional to the ensemble averages of orientation. These early studies form the theoretical framework for our model of homogeneous suspensions of semi-rigid M13 bacteriophage via the approximation of elongated, axisymmetric, rigid Brownian rods in the limit of infinite aspect ratio. Other relevant theoretical studies include work on homogeneous shear \cite{strand1987computation, kamal1989prediction, leal1972rheology}, time-dependent shear \cite{hinch1973time}, turbulent channel flow \cite{manhart2003rheology, marchioli2010orientation, mortensen2008dynamics} and pressure-driven channel flow \cite{ezhilan2015transport}. More recently, Taylor-Couette \cite{holloway2015couette} and Rayleigh-B\'{e}nard \cite{holloway2018linear} stability of perfectly aligned suspensions has been theoretically modelled via the transversely isotropic fluid of Ericksen \cite{ericksen1960ti,green2008extensional}. Similar models have also been used to investigate fibrous biological systems such as growth in plant cell walls \cite{dyson2010fibre}, propulsion in aligned cervical mucus \cite{cupples2017viscous, cupples2018viscous} and the extracellular matrix \cite{dyson2015investigation}.

This paper solves and compares two 3D pressure-driven flow problems for an oriented suspension in a narrow rectangular channel geometry: steady flow, which models the continuously pumped loop used in existing spectroscopic devices, including the LD technology, and a potential novel oscillatory system, in which a smaller volume of analyte is oscillated back and forth. For practical particle concentrations we have $n_d^*{a^*}^3 < 1$, where $n_d^*$ is particle number density and $a^*$ is the particle length, and hence the suspension is at the upper end of the \emph{dilute} range \cite{doi1978dynamics}. The channel dimensions, pressure gradient, particle number density and frequency of oscillations are investigated as factors to improve alignment, and thus signal, in both flow types and to determine the viability of an oscillatory system for aligning particles.

The coupled orientation and flow model will be solved numerically by iterative coupling. Mathematical modelling of these suspensions is computationally challenging due to the additional independent variables associated with the particle degrees of freedom and the coupling between particle dynamics, velocity gradients and rheology. Rational simplification of the flow problem via lubrication theory, the application of a spectral method \cite{strand1987computation} and multicore parallelisation of the array of spatially-local orientation problems, will be shown to enable practical solution with workstation hardware.

The manuscript is organised as follows: the governing equations for the system, including the Navier-Stokes and Fokker-Planck equations, are summarised in section \ref{Sec:Gov}. The steady flow model is presented in section \ref{Sec:Steady} and the oscillatory problem in section \ref{Sec:Osc}. Results for both flow problems are presented in section \ref{Results} and discussed and compared in section \ref{Sec:Disc}.

\begin{figure}
\centering
\includegraphics[scale=0.7, trim=0cm 0cm 0cm 0cm,clip]{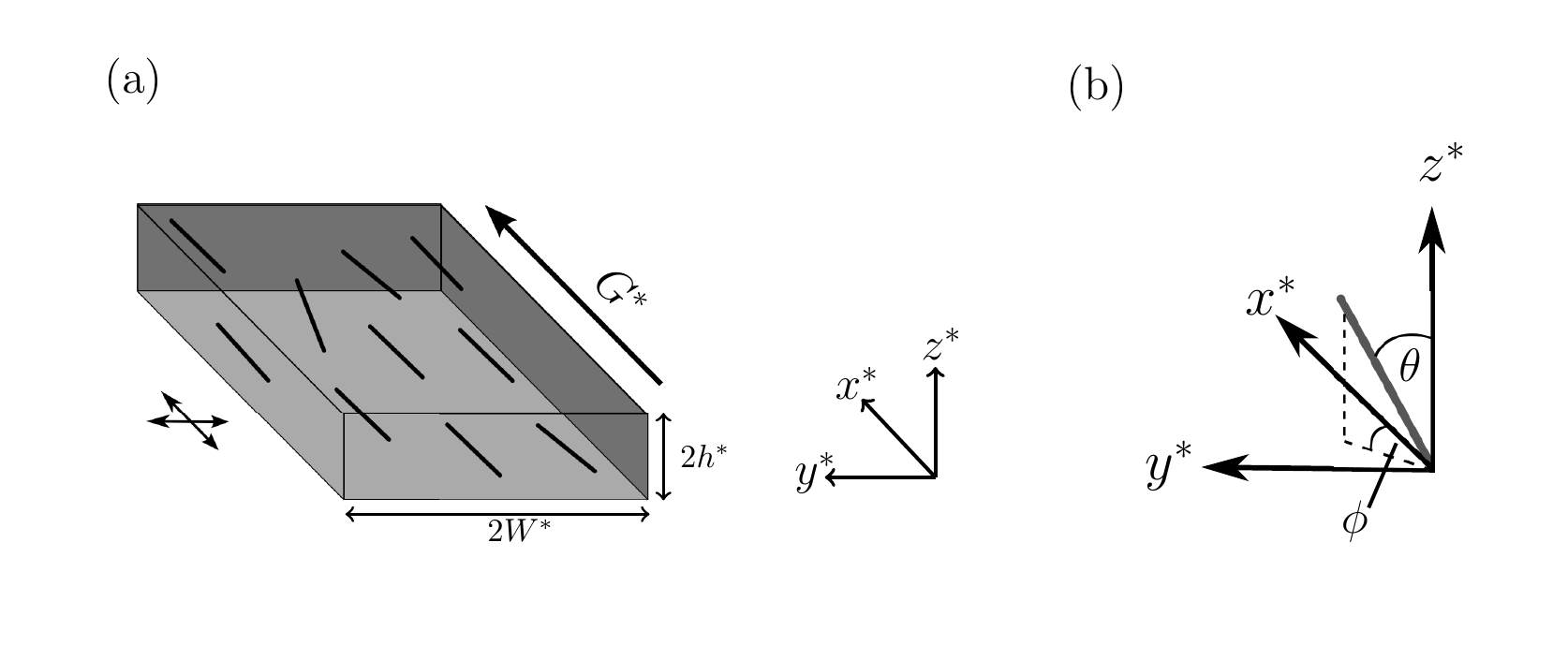}
\caption{Schematic of the flow of a fibre-laden fluid, forced by pressure gradient $G^*$ with a distribution of orientation, in a thin rectangular channel. (a) Rectangular channel of width $2W^*$ and depth $2h^*$ and the $x^*$--axis is the molecular orientation axis. The orientation parameter is calculated as the proportional difference in orientation parallel and perpendicular to the molecular orientation axis, highlighted by the arrows to the left of the channel. Note that for oscillating flow, the pressure gradient is replaced with $G^*e^{i\omega^*t}$ for frequency of oscillations $\omega^*$. (b) The orientation angles $\theta \in [0,\pi]$ and $\phi\in[0,2\pi)$ away from the coordinate axis describe the orientation of the fibres.}
\label{figure1}
\end{figure}

\section{Summary of equations governing dilute suspensions of elongated particles} \label{Sec:Gov}

Consider a 3D rectangular channel of width $2W^*$, depth $2h^*$ and length scale $L^*$, where $h^*,\,W^* \ll L^*$. The axis origin is at the centre of the channel and it is assumed that the flow direction induced by pressure gradient $G^*$, or the molecular orientation axis, is the $x^*$-direction (figure \ref{figure1} (a)). The behaviour of dilute suspensions is approximated by altering the constitutive equation for the stress tensor $\bs{\sigma}^*$ \cite{hinch1975constitutive, hinch1976constitutive}; the motivating problem is at the upper end of the dilute regime (see section \ref{Results}). This constitutive equation, along with the coupled Fokker-Planck equation governing fibre behaviour \cite{strand1987computation}, are briefly detailed now.

The dimensional Navier-Stokes and incompressibility equations are 
\begin{align}
\rho^*\left(\dpd{{\bs{u}^*}}{{t^*}}+(\bs{u}^*\cdot\nabla^*)\bs{u}^*\right)&=\nabla^*\cdot\bs{\sigma}^*, \label{NS}\\
\nabla^*\cdot\bs{u}^*&=0, \label{cont}
\end{align} 
where $\rho^*$ is density, $\bs{u}^*(\bs{x}^*,t^*)$ is the velocity vector for spatial coordinate $\bs{x}^*$ and time $t^*$ and $\nabla^*$ is differentiation in Cartesian space. The asterisk notation represents dimensional variables. The boundary conditions, are given by no-slip and no-penetration at the walls,
\begin{equation} \label{BoundaryConditions}
\bs{u}^*(x^*,\pm W^*,z^*,t^*)=0, \quad \bs{u}^*(x^*,y^*,\pm h^*,t^*)=0.
\end{equation}
The boundary conditions for $x^*$ have not been defined here as they are specified for each flow system.

To determine the bulk stresses in the fluid, consider a homogeneous suspension of particles, dilute enough to ensure particle-particle interactions are negligible \cite{strand1987computation, mclachlan2013calculations}. Further assume the suspension is spatially homogeneous, then the suspension configuration can be described by the orientation distribution function $\psi(\bs{x}^*,\bs{p},t^*)$, satisfying the Fokker-Planck equation,
\begin{equation} \label{FP}
\dpd{\psi}{{t^*} }+\nabla_{\bs{p}}\cdot(\bs{\Omega}^*\psi)=D_r^*\nabla_{\bs{p}}^2\psi,
\end{equation}
where $\bs{p}$ is the unit orientation vector,
\begin{equation}
\bs{p}=(\sin\theta\cos\phi,\sin\theta\sin\phi,\cos\theta),
\end{equation}
for $\theta \in [0,\pi]$ and $\phi \in [0,2\pi)$ shown in figure \ref{figure1} (b), $\nabla_{\bs{p}}$ represents differentiation in orientation space and $D_r^*$ is the rotational diffusion constant. The evolution of the director $\bs{p}$ is given by
\begin{equation} \label{pdot}
\bs{\Omega}^*=(\textbf{I}-\bs{pp})\cdot(\alpha_0\bs{e}^*+\bs{\omega}^*)\cdot\bs{p},
\end{equation}
where $\bs{e}^*=(\nabla^*\bs{u}^*+\nabla^*{\bs{u}^*}^T)/2$ is the rate of strain tensor, $\bs{\omega}^*=(\nabla^*\bs{u}^*-\nabla^*{\bs{u}^*}^T)/2$ is the vorticity tensor and $\alpha_0=(r^2-1)/(r^2+1)$, for particle aspect ratio $r$, is a measure of slenderness \cite{jeffery1922motion}. The orientation distribution function must satisfy periodicity and the normalisation condition,
\begin{align}
\psi(\phi,\theta)&=\psi(\phi+\pi,\theta)=\psi(\phi,\theta+2\pi), \\
& \int_s\psi d\bs{p}=1, \label{norm}
\end{align} 
where $s$ is the unit surface of a sphere.

Following Pedley and Kessler \cite{pedley1990new} (and the prior work of Hinch and Leal \cite{hinch1976constitutive}), the enhanced stress tensor for a suspension of inactive particles is
\begin{equation} \label{extra_stress}
\bs{\sigma}^*=\bs{\sigma}_I^*+\bs{\sigma}_D^*+\bs{\sigma}_P^*,
\end{equation}
where the Newtonian contribution is
\begin{equation} \label{Newtst}
\bs{\sigma}_I^*=-P^*\textbf{I}+2\mu^*\bs{e}^*,
\end{equation}
for pressure $P^*$, identity matrix $\textbf{I}$ and viscosity $\mu^*$.
Here, $\bs{\sigma}_D^*$ is the extra stress due to particle rotation and $\bs{\sigma}_P^*$ represents the interaction of the particles with the fluid, given by
\begin{align} 
\bs{\sigma}_D^*=&2\mu^*\Phi D_r^*\alpha_r\int_s\left(\bs{pp}-\frac{\textbf{I}}{3}\right)\psi\dif \bs{p}, \label{extra_stress_1} \\ 
\begin{split}
\bs\sigma_P^* =& 4 \, \mu^* \, \Phi \Bigg[ \alpha_2 \, {\bs e}^* : \int_s \bs{ p\,p\,p\,p}\, \psi \dif {\bf p} + \alpha_3 \left( {\bs e}^* \cdot \int_s \bs{p \, p}\,\psi \dif {\bs p} + \int_s \bs{p \, p}\, \psi \dif {\bs p} \cdot {\bs e}^* \right) \\
& \hspace{3cm}+ \alpha_4  \, {\bs e}^* \int_s \psi \dif {\bs p} + \alpha_5 \, {\bs I} \,  {\bs e}^* : \int_s \bs{p \, p}\, \psi \dif {\bs p} \Bigg], \label{extra_stress_2}
\end{split}
\end{align}
and where $\Phi$ is the volume fraction of particles. The constants $\alpha_i$ ($i=2\dots5,\,r$) relate to the aspect ratio of the particle and can be represented in terms of ellipsoidal integrals \cite{batchelor1970stress, jeffery1922motion}, which are detailed in equations (S1)--(S5) in supplemental materials section S1. For reference, Holloway et al \cite{holloway2018influences} have summarised this system of equations for particle suspensions in concise notation.

Since the width and depth of the channel are comparable and much smaller than the length scale, we use a lubrication approximation, introducing the dimensionless parameter $\delta=h^*/L^*\ll 1$ to allow detailed analysis of the significant hydrodynamic interactions. The variables are scaled via
\begin{align}  
\begin{split}
u^*=\frac{G^*{h^*}^2}{\mu^*}u, \quad v^*=\frac{\delta G^*{h^*}^2}{\mu^*}v, \quad w^*=\frac{\delta G^*{h^*}^2}{\mu^*}w,\\
\quad x^*=\frac{h^*}{\delta}x, \quad y^*=h^*y, \quad z^*=h^*z, \quad P^*=\frac{G^*h^*}{\delta}P, \label{nondimSCF}
\end{split}
\end{align}
where $G^*$ is a reference pressure gradient. For steady flow, $G^*$ is a complete pressure gradient and for oscillating flow it is determined from the amplitude of the oscillating pressure gradient. The full dimensionless models for steady and oscillatory flow are described in the relevant sections.

Finally, the degree of orientation in a system can be described by an orientation parameter, $S$, calculated as
\begin{align} \label{OP}
\begin{split}
S=&\langle p_{x^*}^2 \rangle - \langle p_{y^*}^2 \rangle, \\
 =&\langle\sin^2\theta\cos^2\phi \rangle - \langle\sin^2\theta\sin^2\phi\rangle,
\end{split}
\end{align}
for a 3D suspension where
\begin{equation}
\langle\bullet\rangle=\int_0^{2\pi}\int_0^{\pi}\bullet\,\psi\sin\theta\dif\theta\dif\phi.
\end{equation}
Note that this is different from the definition used by \cite{mclachlan2013calculations},
\begin{equation} \label{OP_mc}
S=\frac{1}{2}\left(3\int_s\cos^2\theta_S\, \psi \dif \bs{p}-1\right),
\end{equation}
where 
\begin{equation*}
\theta_S=\arccos(|\sin\theta \cos\phi|).
\end{equation*}
The expression \eqref{OP_mc} relates only to uni-axial samples such as planar membranes in which $\psi=\psi(\theta_S,t^*)$ \cite{garab2009linear, norden1997circular}, rather than flow spectroscopy in which the orientation distribution is bi-axial, {\it i.e.} $\psi=\psi(\theta,\phi,t^*)$. McLachlan's\cite{mclachlan2013calculations} use of this formula is therefore inaccurate in this context.

Equations \eqref{NS}--\eqref{extra_stress_2} constitute the full dimensional model for variables $u^*(\bs{x}^*,t^*)$ and $\psi(\phi,\theta,t^*)$ which will be solved numerically. The steady flow system is described in section \ref{Sec:Steady} and the oscillatory system in section \ref{Sec:Osc}.

\section{Steady flow} \label{Sec:Steady}

The first problem we consider is flow forced by a constant pressure gradient $G^*$. Section \ref{Sec:Steady} \ref{Sec:FP} contains the Fokker-Planck equation and section \ref{Sec:Steady} \ref{Sec:SteadyVel} contains the model for the flow. The full coupled problem is solved by iterating between the numerical methods described in section \ref{Sec:Steady} \ref{Sec:FP} and section \ref{Sec:Steady} \ref{Sec:SteadyVel}.

\subsection{Model for the orientation distribution function} \label{Sec:FP}

The particles in the flow are small enough that each particle at a point in space is subject to an approximately linear velocity field in the local frame, and hence a constant shear rate $\dot{\gamma}^*=\sqrt{(\nabla^*\bs{u}^*)^2+(\nabla^*{\bs{u}^*}^T)^2}$. Therefore the orientation distribution of the bacteriophage is formulated at each point in space independently; fluid flow influences orientation distribution only via the local P\'{e}clet number. We describe the model and its numerical discretisation below. Following Strand, Kim and Karrila \cite{strand1987computation} (see also Bird et al \cite{bird1977dynamics}), assuming $\alpha_0=1$ and investigating steady state behaviour, the Fokker-Planck equation reduces to
\begin{equation} \label{FP_SKKstead}
\frac{1}{6}\Lambda(\psi)=P_e\Upsilon(\psi),
\end{equation}
where $P_e=\dot{\gamma}^*/6D_r^*$ is the local P\'{e}clet number and
\begin{align}
\Lambda(\psi)&=\dfrac{1}{\sin{\theta'}}\dpd{}{{\theta'}}\left(\sin{\theta'}\dpd{\psi}{{\theta'}}\right)+\dfrac{1}{\sin^2{\theta'}}\dpd[2]{\psi}{{\phi'}}, \label{Lambda} \\ 
\Upsilon(\psi)&=\dfrac{\sin{\phi'}\cos{\phi'}}{\sin{\theta'}}\dpd{}{{\theta'}}\left(\sin^2{\theta'}\cos{\theta'}\,\psi\right)-\dpd{}{{\phi'}}\left(\sin^2{\phi'}\,\psi\right), \label{Upsilon}
\end{align}
where $\theta'$ and $\phi'$ describe the local coordinates at each point in space, which are calculated by rotating $\phi$ and $\theta$ away from the lab frame by an angle $\beta=\text{acos}(|\partial u^*/\partial y^*| / \dot{\gamma}^*)$.

The Fokker-Planck equation \eqref{FP_SKKstead} is spatially discretised via spherical harmonics \cite{bird1977dynamics, mclachlan2013calculations, strand1987computation}. Assume the solution takes the form
\begin{equation} \label{SpherHarStead}
\psi({\phi'},{\theta'})=\sum_{n=0}^N\sum_{m=0}^n\left(A_{0\,n}^m P_n^m(\cos{\theta'})\cos(m{\phi'})+A_{1\,n}^m P_n^m(\cos{\theta'})\sin(m{\phi'})\right),
\end{equation}
where $P_n^m$ are the associated Legendre polynomials. By substituting \eqref{SpherHarStead} into equation \eqref{FP_SKKstead}, $\Lambda$ and $\Upsilon$ are given by
\begin{align}
\Lambda(P_n^m(\cos{\theta'})\cos(m{\phi'}))&=-n(n+1)P_n^m(\cos\theta')\cos(m\phi'), \label{L1}\\
\Lambda(P_n^m(\cos{\theta'})\sin(m{\phi'}))&=-n(n+1)P_n^m(\cos\theta')\sin(m\phi'), \label{L2}\\
\Upsilon(P_n^m(\cos{\theta'})\cos(m{\phi'}))&=-\sum_{p=m-2}^{m+2}\sum_{q=n-2}^{n+2}a_{n,q}^{m,p}P_n^m(\cos{\theta'})\sin(m{\phi'}), \quad m\geq 0, \label{O1} \\
\Upsilon(P_n^m(\cos{\theta'})\sin(m{\phi'}))&=\sum_{p=m-2}^{m+2}\sum_{q=n-2}^{n+2}a_{n,q}^{m,p}P_n^m(\cos{\theta'})\cos(m{\phi'}), \quad m > 0. \label{O2}
\end{align}
The first two results follow from the definition of a spherical harmonic, {\it i.e.} the solution to the spherical part of Laplace's equation on a sphere; the other two are a linear combination of spherical harmonics where the seven non-zero constants $a_{n,q}^{m,p}$, reproduced from Bird \cite{bird1977dynamics} and Strand, Kim \& Karrila \cite{strand1987computation}, are given in supplemental materials section S2. The resulting system of equations for $A_{0\,n}^m$ and $A_{1\,n}^m$, found by equating coefficients of sine and cosine, is
\begin{align}
\frac{q(q+1)}{6}A_{0\,q}^p&=-P_e\sum_{n=0}^{N}\sum_{m=0}^{n}a_{n,q}^{m,p}A_{1\,n}^m, \label{eqA0stead}\\
\frac{q(q+1)}{6}A_{1\,q}^p&=P_e\sum_{n=0}^{N}\sum_{m=0}^{n}a_{n,q}^{m,p}A_{0\,n}^m, \label{eqA1stead}
\end{align}
for integer $N$, $q=0,2,\ldots,N$ and $p=0,2,\ldots, q$. Here, $A_{0\,0}^0=1/4\pi$ to satisfy the normalisation condition, and we have $A_{1\,n}^0=0$ for all $n$. Due to particle symmetry, if $n$ or $m$ are odd then $A_{0\,n}^m=A_{1\,n}^m=0$.

The system \eqref{eqA0stead}--\eqref{eqA1stead} is solved numerically by setting up a matrix of equations for $A_{0\,q}^p$ and $A_{1\,q}^p$ and using a direct solver (backslash) in Matlab.

\subsection{Model for the velocity} \label{Sec:SteadyVel}

Under the lubrication theory approximation, the flow problem can be expressed entirely as an elliptic partial differential equation for $u(y,z)$ where the coefficients depend on moments of the orientation distribution $\psi(\phi,\theta,y^*,z^*)$.

Scaling the system, via \eqref{nondimSCF}, neglecting terms $\mathcal{O}(\delta)$ and smaller, the continuity equation \eqref{cont} is unchanged and the Navier-Stokes equations \eqref{NS} reduce to

\begin{align}
\begin{split}
-1=&\dpd{}{y}\left(\left\{1+4\Phi\left[\alpha_2\int_{s}p_1^2p_2^2\psi\,\dif\bs{p}+\dfrac{\alpha_3}{2}\int_{s}(p_1^2+p_2^2)\psi\,\dif\bs{p}+\frac{\alpha_4}{2} \right] \right\}\dpd{u}{y} \right.  \\
  &\hspace{3.6cm} \left. +4\Phi\left[\alpha_2 \int_{s}p_1^2p_2p_3\psi\,\dif\bs{p} +\dfrac{\alpha_3}{2}\int_{s}p_2p_3\psi\,\dif\bs{p}\right]\dpd{u}{z} \right) \\
 +&\dpd{}{z}\left(\left\{1+4\Phi\left[\alpha_2\int_{s}p_1^2p_3^2\psi\,\dif\bs{p}+\dfrac{\alpha_3}{2}\int_{s}(p_1^2+p_3^2)\psi\,\dif\bs{p}+\frac{\alpha_4}{2} \right] \right\}\dpd{u}{z} \right. \\
  &\hspace{3.6cm}\left. +4\Phi\left[\alpha_2 \int_{s}p_1^2p_2p_3\psi\,\dif\bs{p} +\dfrac{\alpha_3}{2}\int_{s}p_2p_3\psi\,\dif\bs{p}\right]\dpd{u}{y} \right) \\
 +&\dfrac{2\Phi\alpha_r}{P_G}\left(\dpd{}{y}\int_{s}p_1p_2\psi\,\dif\bs{p}+\dpd{}{z}\int_{s}p_1p_3\psi\,\dif\bs{p}\right), \label{SteadyGov}
\end{split}
\end{align}
for global P\'{e}clet number $P_G=G^*h^*/D_r^*\mu^*$. The $y$-- and $z$--components of the dimensionless Navier-Stokes equations enforce $P=P(x)$ and so the pressure gradient reduces to $-1$. The boundary conditions \eqref{BoundaryConditions} become
\begin{equation} \label{dimless_bc}
u(\pm\overline{W},z)=0, \qquad u(y,\pm 1)=0,
\end{equation}
where $\overline{W}=W^*/h^*$. Equation \eqref{SteadyGov}, along with boundary conditions \eqref{dimless_bc}, are solved numerically using a finite difference scheme and constructing a matrix of equations from the resulting problem. The numerical solution to the Newtonian and suspension models are described in supplemental materials section S3 and the iterative process is described in section S4.

To optimise the efficiency of the above processes, the convergence of the spherical harmonic solution and the finite difference approximation are investigated for a range of spherical harmonic modes $N$ and finite difference grid spacings $Y$ and $Z$. The full details and related figures are presented in supplemental materials section S3; we find that $N=10$, $Y=150$ and $Z=120$ are sufficient for convergence. We will discuss the results in section \ref{Results}.

\section{Oscillatory flow} \label{Sec:Osc}

The next flow system of interest consists of an oscillating pressure gradient where the length is still much larger than its width and depth but a smaller volume of fluid oscillates back and forth. The pressure gradient is now defined as
\begin{equation} \label{oscG}
\dpd{{P^*}}{{x^*}}=G^*\exp(i\,\omega^*t^*),
\end{equation}
for frequency of oscillations $\omega^*$. The dimensional scalings are given by equation \eqref{nondimSCF}, and the time is scaled against the frequency of oscillations as $t^*=t/\omega^*$. 

The model for the time dependent Fokker-Planck equation is given in section \ref{Sec:Osc} \ref{Sec:FPOsc} and the model for the velocity profile is given in section \ref{Sec:Osc} \ref{Sec:VelOsc}. The full coupled problem comes from iterating numerically between \ref{Sec:Osc} \ref{Sec:FPOsc} and \ref{Sec:Osc} \ref{Sec:VelOsc} over each time step.

\subsection{Model for the orientation distribution} \label{Sec:FPOsc}

Assuming the particles are small enough that at each point in the channel they are subjected to a spatially linear velocity field and thus a shear rate dependent only on time $\dot{\gamma}^*=\dot{\gamma}^*(t^*)$; the orientation distribution is again formulated at each point in space independently. Following section \ref{Sec:Steady} \ref{Sec:FP}, the Fokker-Planck equation for shear rate $\dot{\gamma}(t)$ is 
\begin{equation} \label{FP_SKKosc}
\dpd{\psi}{\tau}=\frac{1}{6}\Lambda(\psi)-P_e(\tau)\Upsilon(\psi),
\end{equation}
where $P_e=\dot{\gamma}^*/6D_r^*$ is the local P\'{e}clet number, $t^*=\tau/6D_r^*$ and $\Lambda$ and $\Upsilon$ are given by \eqref{Lambda} and \eqref{Upsilon} respectively.

The orientation distribution is discretised spatially in the basis of spherical harmonics,
\begin{equation} \label{SpherHarosc}
\psi(\phi',\theta',t)=\sum_{n=0}^N\sum_{m=0}^n\left(A_{0\,n}^m(t)P_n^m(\cos\theta')\cos(m\phi')+A_{1\,n}^m(t)P_n^m(\cos\theta')\sin(m\phi')\right).
\end{equation}
This is then substituted into \eqref{FP_SKKosc}, where $\Lambda$ and $\Upsilon$ are given by \eqref{L1}-\eqref{O2}. Equating coefficients of sine and cosine, the resulting system of ordinary differential equations for $A_{0\,n}^m$ and $A_{1\,n}^m$ is
\begin{align}
\dod{A_{0\,q}^p}{t}&=-\dfrac{q(q+1)}{6}A_{0\,q}^p-P_e(t) \displaystyle\sum_{n=0}^{N}\sum_{m=0}^{n}a_{n,q}^{m,p}A_{1\,n}^m, \label{eqA0osc}\\
\dod{A_{1\,q}^p}{t}&=-\dfrac{q(q+1)}{6}A_{1\,q}^p+P_e(t) \displaystyle\sum_{n=0}^{N}\sum_{m=0}^{n}a_{n,q}^{m,p}A_{0\,n}^m, \label{eqA1osc}
\end{align}
for integer $N$, $q=0,2,\ldots,N$ and $p=0,2,\ldots, q$. Here, $A_{0\,0}^0(t)=1/4\pi$ to satisfy the normalisation condition, $A_{1\,n}^0=0$ for all $n$ and $A_{0\,n}^m=A_{1\,n}^m=0$ for odd $n,\,m$. 

The ordinary differential equations \eqref{eqA0osc} and \eqref{eqA1osc} are solved numerically using a second order explicit Improved Euler method by first writing the equations in matrix form as described in section \ref{Sec:Steady} \ref{Sec:FP}.


\subsection{Model for the velocity} \label{Sec:VelOsc}

Using a lubrication theory approximation, the velocity problem can be expressed as a time dependent partial differential equation where coefficients of $u(y,z,t)$ depend upon moments of the orientation distribution function $\psi(\phi,\theta,y^*,z^*,t^*)$.

Scaling the Navier-Stokes and continuity equations, \eqref{NS} and \eqref{cont} respectively, retaining terms larger than $\mathcal{O}(\delta)$, the continuity equation is unchanged and the dimensionless Navier-Stokes equations become
\begin{align}
\begin{split} 
\alpha^2\dpd{u}{t}=&-\exp(it)+\dfrac{2\Phi\alpha_r}{P_G}\left(\dpd{}{y}\int_{s}p_1p_2\psi\,\dif\bs{p}+\dpd{}{z}\int_{s}p_1p_3\psi\,\dif\bs{p}\right) \\
+&\dpd{}{y}\left(\left\{1+4\Phi\left[\alpha_2\int_{s}p_1^2p_2^2\psi\,\dif\bs{p}+\dfrac{\alpha_3}{2}\int_{s}(p_1^2+p_2^2)\psi\,\dif\bs{p}+\dfrac{\alpha_4}{2} \right] \right\}\dpd{u}{y} \right. \\
 & \hspace{2.5cm} \left. +4\Phi\left[\alpha_2 \int_{s}p_1^2p_2p_3\psi\,\dif\bs{p} +\dfrac{\alpha_3}{2}\int_{s}p_2p_3\psi\,\dif\bs{p}\right]\dpd{u}{z} \right) \\
 +&\dpd{}{z}\left(\left\{1+4\Phi\left[\alpha_2\int_{s}p_1^2p_3^2\psi\,\dif\bs{p}+\dfrac{\alpha_3}{2}\int_{s}(p_1^2+p_3^2)\psi\,\dif\bs{p}+\dfrac{\alpha_4}{2} \right] \right\}\dpd{u}{z} \right. \\
&  \hspace{2.5cm}\left. +4\Phi\left[\alpha_2 \int_{s}p_1^2p_2p_3\psi\,\dif\bs{p} +\dfrac{\alpha_3}{2}\int_{s}p_2p_3\psi\,\dif\bs{p}\right]\dpd{u}{y} \right), \label{OscillatingGov}
\end{split}
\end{align}
for global P\'{e}clet number $P_G$ and Womersley number $\alpha^2=\omega^*{h^*}^2\rho^*/\mu^*$. It is assumed the system is initially at rest, accounted for by altering the dimensionless pressure gradient to $\partial P/ \partial x=\exp(i(t+\pi/2))$. 

Equation \eqref{OscillatingGov} is solved via an Alternating Direction Implicit (ADI) method in which the time step is split in half and one spatial variable is discretised in each time step \cite{douglas1955numerical}; mixed derivatives are treated explicitly throughout this analysis. As in section \ref{Sec:Steady}, a matrix equation is constructed for each time step and solved numerically; the details of this method can be found in supplemental materials section S5 and a description of the iterative procedure is described in S6.

A numerical convergence study was undertaken, detailed in supplemental materials section S5. As in section \ref{Sec:Steady}, $Y=150$ and $Z=120$ are satisfactory in the finite difference scheme and $T=1200$ time steps ensure convergence.

\section{Results} \label{Results}

The velocity, orientation parameter and linear dichroism signal are compared for a range of particle number density, channel dimensions and pressure gradient. We will refer to the standard parameter set for these quantities as: channel half-depth $h^*=3\times10^{-4}~$m, half-width $W^*=1\times10^{-3}~$m, fixed pressure gradient $G^*\approx4.4\times 10^3~$Pa/m, particle number density $n_d^*=7.33\times10^{17}~$phage/m$^3$ and frequency of oscillations $\omega^*=10~$s$^{-1}$ (table \ref{param} (a)). Using a particle length of $a^*=800~$nm, the value $n_d^*{a^*}^3=0.3753<1$ and so we argue that the approximation of a dilute regime is reasonable. The dimensionless groups are detailed in table \ref{param} (c) and are subject to change depending on their composition of dimensional values; we do not directly vary the dimensionless groups. All other parameters are unchanged (table \ref{param} (b)). 


\begin{table} 
\begin{center}
\vspace{0.3cm}
(a) Variable parameters \\
\begin{tabular}{ c  c  c  c  c }
\hline
Parameter & Description & Standard value & Range & Units 	\\ [0.5ex]
\hline
$h^*$ & Channel half-depth & $3\times 10^{-4}$ & $2\times10^{-4}$ -- $4\times10^{-3}$ & m  \\[0.7ex]
$W^*$ & Channel half-width & $1\times 10^{-3}$ & $0.4\times10^{-3}$ -- $2.5\times10^{-3}$   & m \\[0.7ex]
$n_d^*$ & Number density & $7.33\times 10^{17}$ & $1\times10^{16}$ -- $1\times10^{18}$ & phage/m$^3$ \\[0.7ex]
$G^*$ & Pressure gradient & $4.4\times 10^3$ & $0.4\times10^3$ -- $10\times10^{3}$ & Pa/m \\[0.7ex]
$\omega^*$ & Frequency of oscillations & $10$ & $5$ -- $70$ & s$^{-1}$ \\[0.7ex]
\end{tabular}\\
\vspace{0.5cm}
(b) Fixed parameters and physical constants\\
\begin{tabular}{ c  c  c  c }
\hline
Parameter  & Description	& Value & Units \\ [0.5ex]
\hline
$a^*$  	   & Particle length & $8\times 10^{-7}$ & m  \\[0.7ex]
$b^*$    	   & Particle width  & $6\times 10^{-9}$ & m  \\[0.7ex]
$V_c^*$  	   & Particle volume & $1.21\times 10^{-21}$  & m$^{3}$ \\[0.7ex]
$T^*$    	   & Temperature & $295$  & K  \\[0.7ex]
$\mu^*$  	   & Viscosity (water) & $9.5\times 10^{-4}$  & Pa s 	 \\[0.7ex]
$\rho^*$ 	   & Fluid density (water)& $1\times 10^3$  & kg/m$^3$\\[0.7ex]
$D_r^*$  	   & Rotational diffusion coefficient & $40.65$ & s$^{-1}$ \\[0.7ex]
$m_w^*$      & Particle molecular weight & $2\times10^7$ & g/mol \\[0.7ex]
$\epsilon^*$ & Extinction coefficient & $0.38$ & m$^2$/g  \\[0.7ex]
$K_B^*$  	   & Boltzmann constant & $1.38\times 10^{-23}$ & kg m$^2$/K s$^2$ \\[0.7ex]
$N_A^*$ 	   & Avogadro constant & $6\times10^{23}$ & phage/mol  \\[0.7ex]
\end{tabular}\\
\vspace{0.5cm}
(c) Dimensionless groups \\
\begin{tabular}{ c  c  c  c }
\hline
Parameter & Description & Definition & Value \\ [0.5ex]
\hline
$\Phi$     & Volume fraction & $V_c^*n_d^*$ & $1.11\times 10^{-5}$ \\[0.7ex]
$P_G$	 & P\'{e}clet number (global) & $G^*h^*/D_r^*\mu^*$ & $34.17$ \\[0.7ex]
$\alpha^2$ & Womersley number & $\omega^*{h^*}^2\rho^*/\mu^*$ & $18.95$ \\[0.7ex]
\end{tabular}
\end{center}
\caption{Summary of the standard parameter set split into three categories: (a) the parameters that vary throughout the results, (b) the fixed parameters and physical constants and (c) the dimensionless groups.}
\label{param}
\end{table}

The notation $\overline{\,\bullet\,}$ and $\overline{\,\bullet\,}^{y^*}$ represent the spatial average and the width average for steady flow respectively,
\begin{equation} \label{spatialAv}
\overline{\,\bullet\,}=\frac{1}{4W^*h^*}\int_{-W^*}^{W^*}\int_{-h^*}^{h^*}\bullet \dif y^* \dif z^*, \quad
\overline{\,\bullet\,}^{y^*}=\frac{1}{4W^*}\int_{-W^*}^{W^*}\bullet \dif y^*.
\end{equation} 
The linear dichroism signal is assumed proportional to the particle number density and is calculated as the root mean square of the orientation parameter across the depth of the channel. The root mean square of the orientation parameter will be denoted as $S_{rms}$ and the signal is given by 
\begin{equation}
LD=\frac{\epsilon^* m_w^* n_d^*}{N_A^*}S_{rms},
\end{equation} 
where $N_A^*$ is the Avogadro constant ($\approx 6 \times 10^{23}$ phage/mol), $m_w^*$ is the molecular weight of M13 bacteriophage ($m_w^*\approx2\times 10^7~$g/mol from Linear Diagnostics Ltd) and $\epsilon^*$ is the extinction coefficient (m$^2$/mol), which is a measure of the absorption of light through the sample \cite{daviter2013circular}. The extinction coefficient is taken as $0.38~$m$^2$/g at light wavelength $269~$nm \cite{aubrey1991raman, glucksman1992three, niu2008bacteriophage}. A number of the parameters in table \ref{param} and their calculation are discussed in supplemental materials section S7. As a direct comparison with steady flow, the maximum value of the orientation parameter in the oscillating channel is presented.

The computational walltime for this problem was large, especially for oscillating flow. The numerical solution was parallelised using {\it `parfor'} in Matlab by splitting the solution to the Fokker-Planck equation, solved at each point in space, across five cores on a workstation (2017 Lenovo Thinkstation P710; Intel(R) Xeon(TM) E5-2646 CPU @ 2.40 GHz; 128 GB 2400 MHz RDIMM RAM). The runtime varied between $14$ and $20$ hours for a single parameter tuple; $175$ tuples were calculated for the oscillating results.

\subsection{Steady pressure gradient}

\begin{figure}[!h]
\centering
\includegraphics[scale=0.6, trim=1.15cm 0.4cm 0cm 0.7cm,clip]{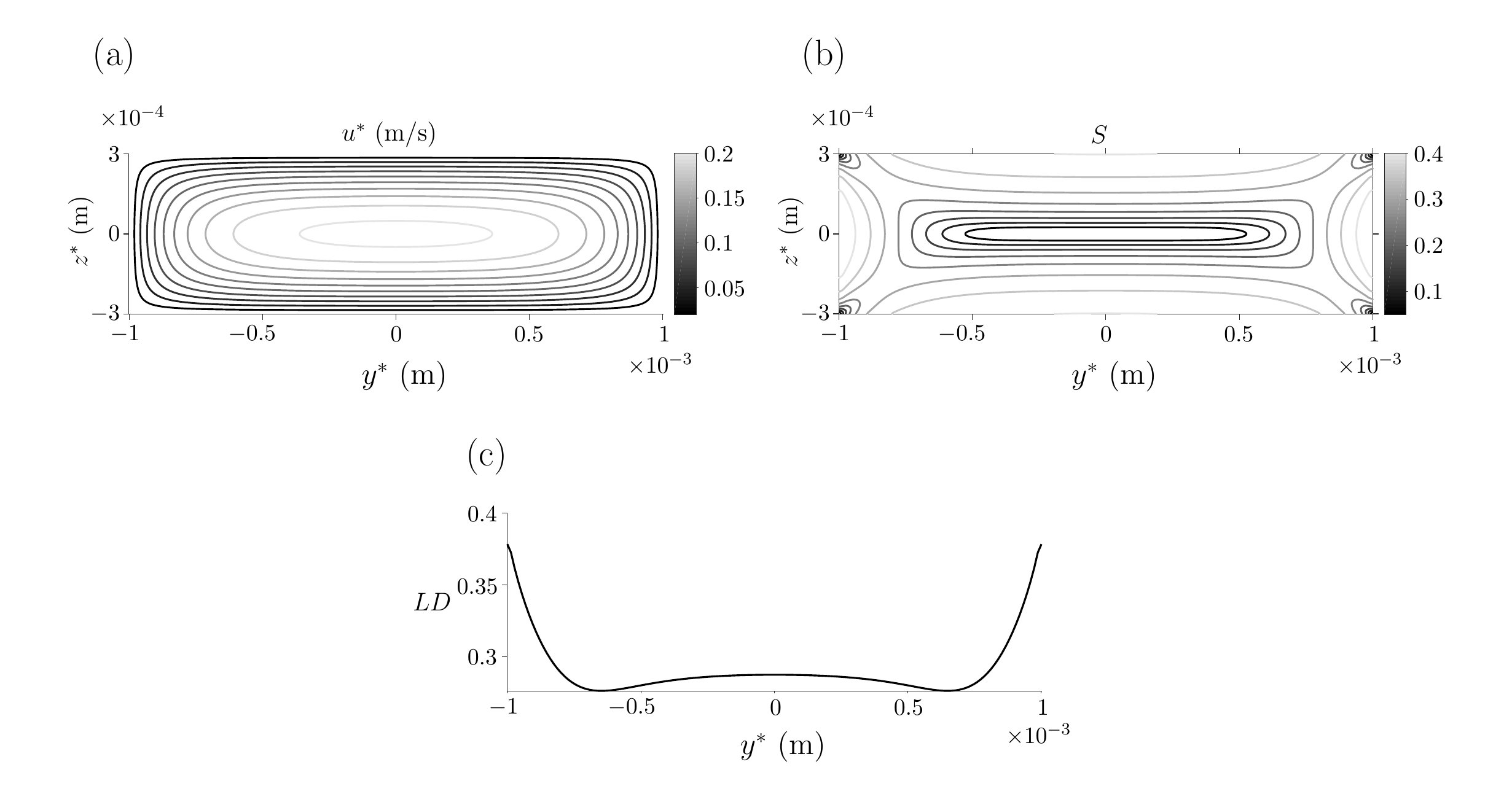}
\caption{Velocity field, orientation parameter and predicted linear dichroism signal for steady flow and the standard parameter set. (a) Velocity profile, (b) orientation parameter and (c) linear dichroism signal.}
\label{figure2}
\end{figure}

Figure \ref{figure2} depicts the velocity profile, orientation parameter and linear dichroism signal for steady flow through the cross section of the channel, with the standard parameter set. As may be expected from the Newtonian solution\cite{pozrikidis2011fluid}, the velocity $u^*$ increases from zero at the channel walls to a maximum at the centre point (figure \ref{figure2} (a)), corresponding to a minimum in the orientation parameter $S$ (figure \ref{figure2} (b)). Associated with the large shear rate, the linear dichroism signal is largest near the channel walls and decreases to a local minimum at the centre (figure \ref{figure2} (c)); there is an increase at the middle width of the channel ($y^*=0$) where larger shear near the walls at $z^*=\pm h^*$ increases the orientation parameter. Local minima also exist at the channel corners.

\begin{figure}[!h]
\centering
\includegraphics[scale=0.72, trim=1.1cm 1.3cm 0cm 0.8cm,clip]{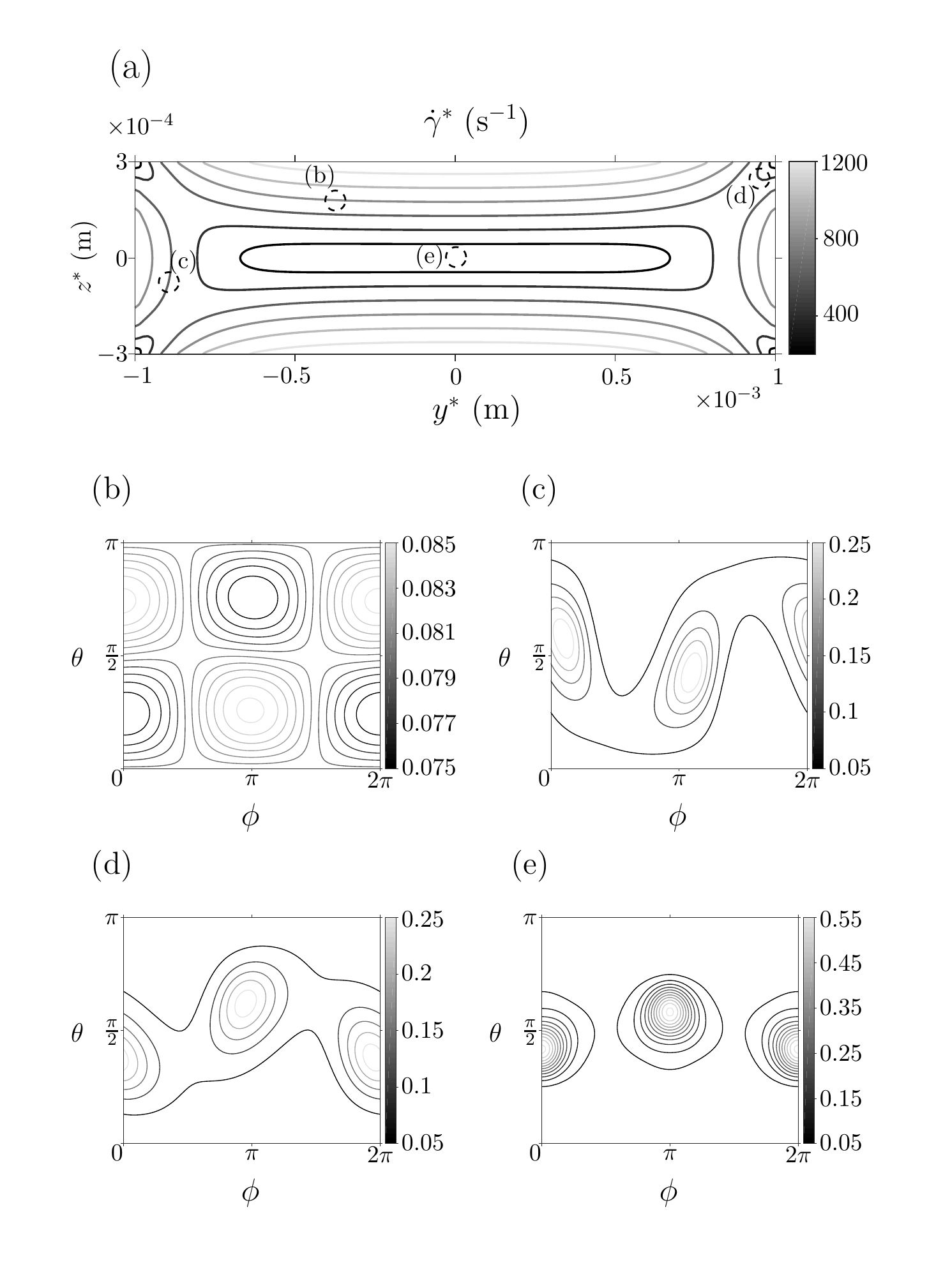}
\caption{The orientation distribution function $\psi$, in orientation space $\psi \in [0,\, 2\pi]$, $\theta \in [0,\, \pi)$, at different points in the cross sectional face of the flow using the standard parameter set. (a) The shear rate $\dot{\gamma}^*$ in the cross sectional face of the channel. The dashed circles, with corresponding labels, show the points in the flow where $\psi$ has been evaluated: (b) $y^*=-0.8\times 10^{-3}~$m, $z^*=-1\times10^{-4}~$m, (c) $y^*=-0.4\times 10^{-3}~$m, $z^*=2\times10^{-4}~$m, (d) $y^*=0.95\times 10^{-3}~$m, $z^*=2.75\times10^{-4}~$m and (e) $y^*=0~$m, $z^*=0~$m.}
\label{figure3}
\end{figure}

The orientation distribution function $\psi$ is displayed for varying points in the channel cross section (figure \ref{figure3}).  At the centre of the channel where the shear rate is zero (figure \ref{figure3} (b)) there are no biasing effects on the particles and so $\psi$ is small. Close to the walls at $y^*=\pm W^*$m the behaviour of $\psi$ is similar (figures \ref{figure3} (c) and (d)); the particles prefer to align in the flow direction, for orientation angles $\phi=0,\,\pi$ and $2\pi$ and $\theta=\pi/2$.  Close to the corner of the channel this behaviour is more prominent. Near $z^*=h^*$m (figure \ref{figure3} (e)), the M13 bacteriophage are more likely to be aligned at $\phi=0$ and $\pi$ (note that the particles have no polarity so the two angles are equivalent) and $\theta=\pi/2$.

\begin{figure}[!h]
\centering
\includegraphics[scale=0.72, trim=1cm 1.4cm 0cm 0.8cm,clip]{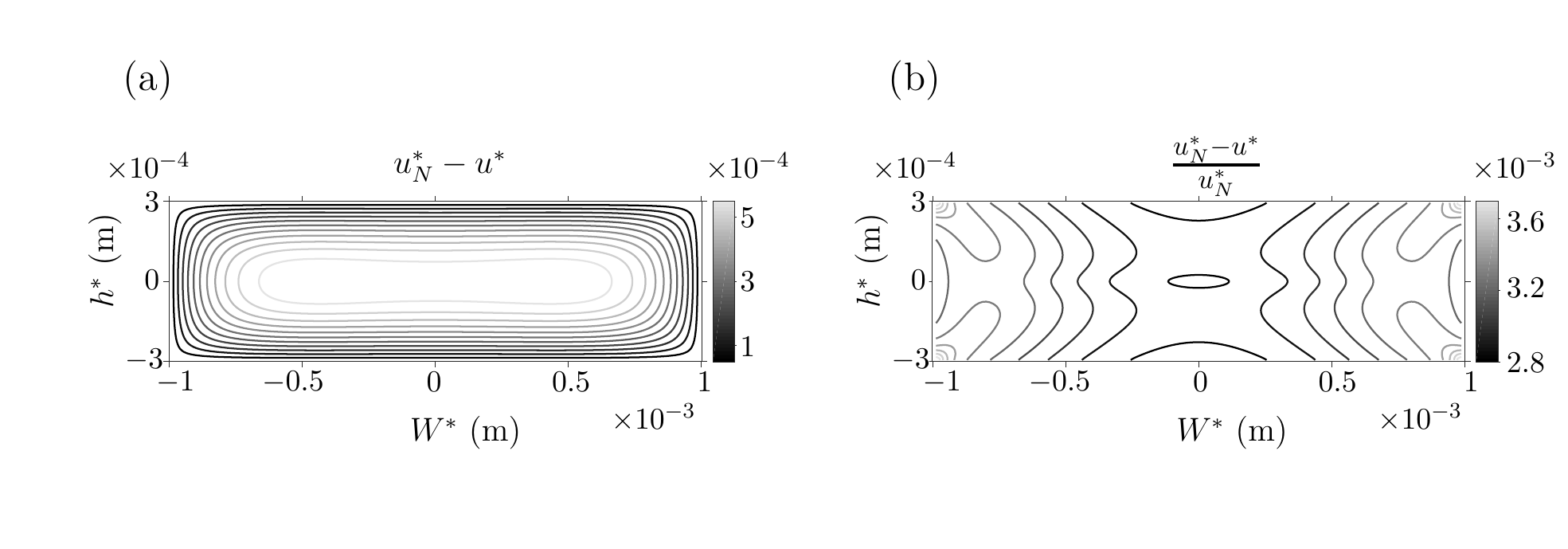}
\caption{The difference between the Newtonian and suspension flow fields in the cross-sectional face of the channel. (a) Depicts the actual difference $u_{N}^*-u^*$ and (b) depicts the difference relative to the magnitude of the Newtonian velocity $(u_{N}^*-u^*)/u_{N}^*$.}
\label{figure4}
\end{figure}

The flow profile accounting for a particle suspension is compared with the Newtonian profile $u_{N}^*$ for the same flow setup in figure \ref{figure4}. The actual difference shows that the profile for both flow systems is qualitatively the same; the suspension model produces a lower velocity however the change is small (figure \ref{figure4} (a)). To confirm that this change is small, the reduction in velocity relative to the magnitude of the Newtonian velocity is also calculated in figure \ref{figure4} (b) and the change is on the order of $10^{-3}$.

\begin{figure}[!h]
\centering
\includegraphics[scale=0.67, trim=1.1cm 0.3cm 0cm 0.5cm,clip]{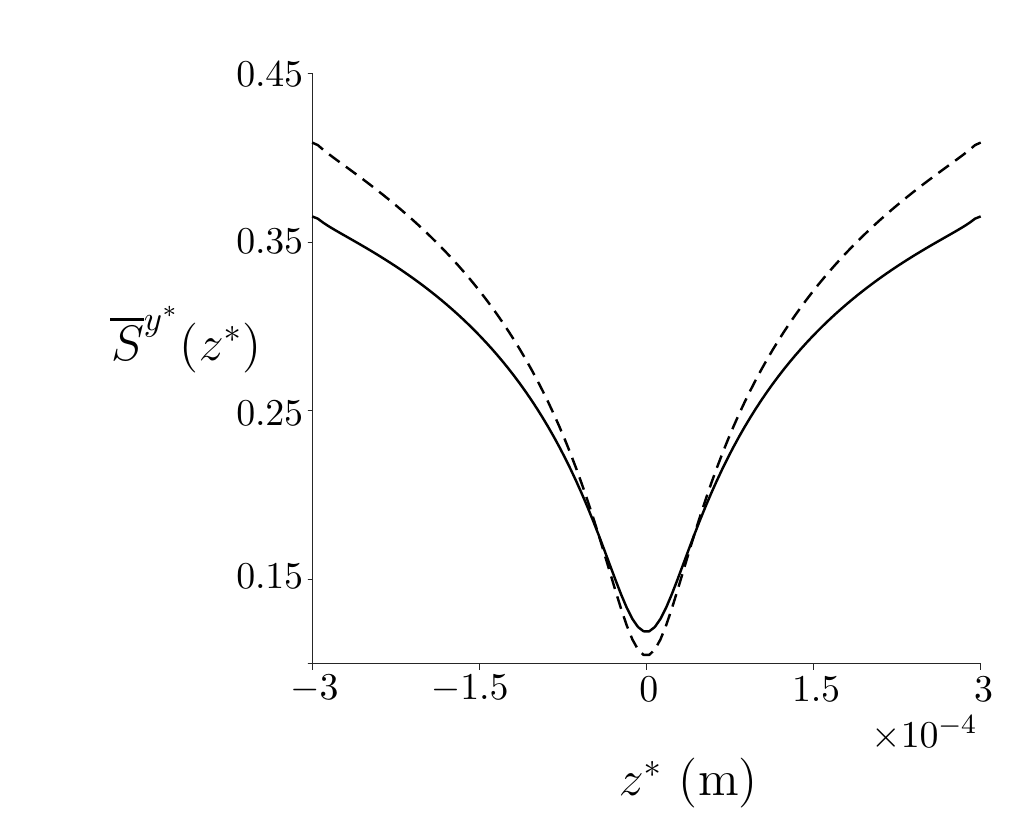}
\caption{A comparison between the width averaged orientation parameter calculated in the current work (solid line) and the definition used by McLachlan \cite{mclachlan2013calculations} (dashed line), for the standard parameter set.}
\label{figure5}
\end{figure}
 
We compare the width averaged orientation parameter calculated in this work (solid line) with equation \eqref{OP_mc} (dashed line); while both definitions show maxima and minima of the orientation parameter at the same point in the channel, McLachlan's definition for the orientation parameter \cite{mclachlan2013calculations} over-predicts the orientation in general.

\begin{figure}[!h]
\centering
\includegraphics[scale=0.63, trim=1.1cm 0.4cm 0cm 0.7cm,clip]{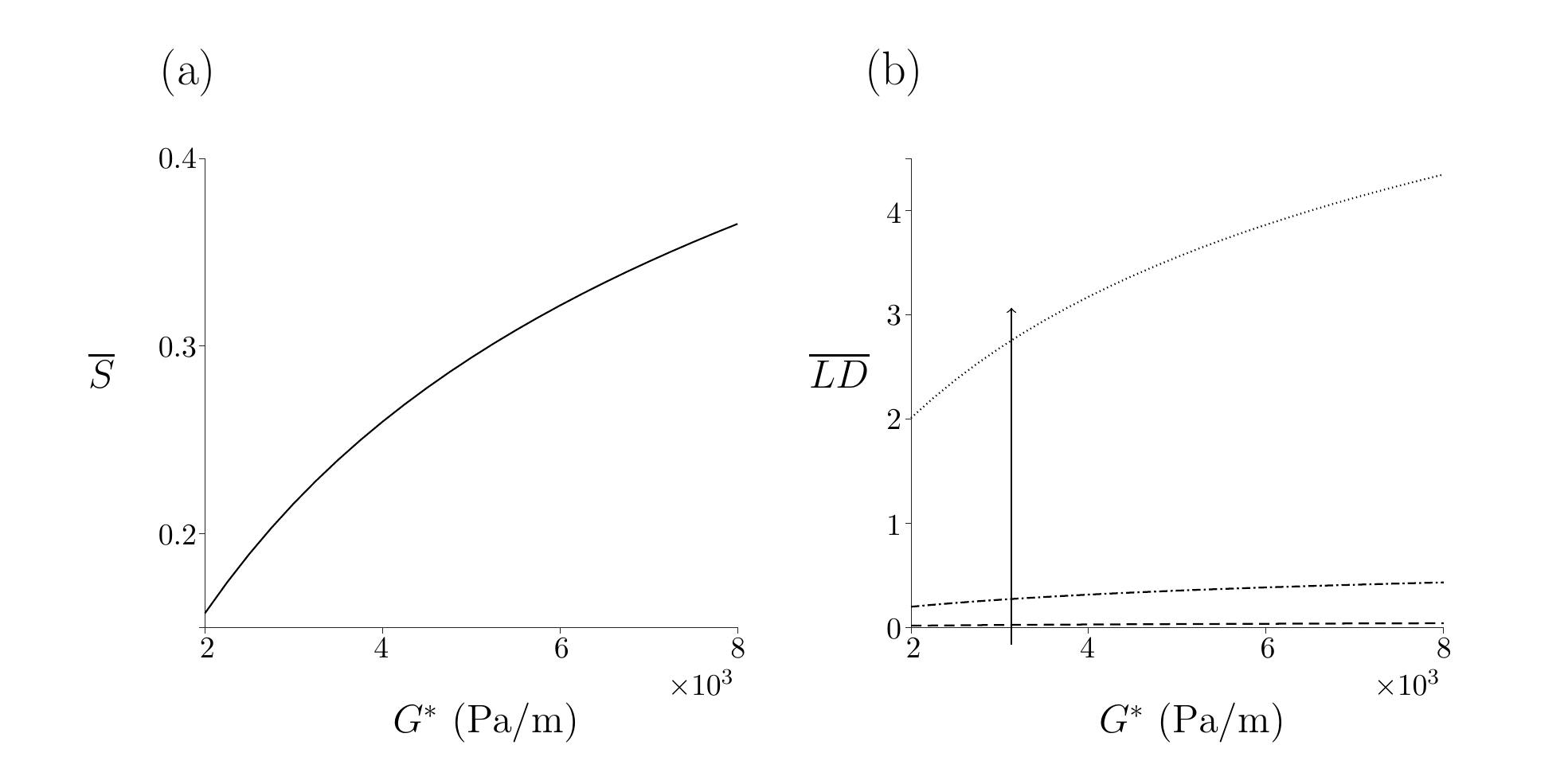}
\caption{The orientation parameter and linear dichroism signal for steady flow with increasing pressure gradient number density. (a) The orientation parameter $\overline{S}$ with increasing pressure gradient, where $n_d^*=1\times10^{18}~$phage/m$^{3}$. (b) The linear dichrosim signal $\overline{LD}$ with increasing pressure gradient and three different particle number densities:  $n_d^*=1\times10^{16}~$phage/m$^3$ (dashed line), $n_d^*=1\times10^{17}~$phage/m$^3$ (dot-dashed line) and $n_d^*=1\times10^{18}~$phage/m$^3$ (dotted line); the arrow denotes increasing number density. The other parameters are $h^*=3\times 10^{-4}~$m and $W^*=1\times 10^{-3}~$m.}
\label{figure6}
\end{figure}

In figure \ref{figure6}, the impact of the pressure gradient $G^*$ and particle number density $n_d^*$ on the spatially averaged orientation parameter $\overline{S}$ and linear dichroism signal $\overline{LD}$ is examined. Increasing the number density reduces the orientation parameter less than $0.5\%$ over two orders of magnitude (results not shown here) and increasing the pressure gradient produces a (non-linear) increase in $\overline{S}$, seen in figure \ref{figure6} (a). However, increasing the number density significantly increases the linear dichroism signal due to its linear relationship with number density (figure \ref{figure6} (b)).

\begin{figure}[!h]
\centering
\includegraphics[scale=0.62, trim=1.1cm 0.6cm 0cm 0.7cm,clip]{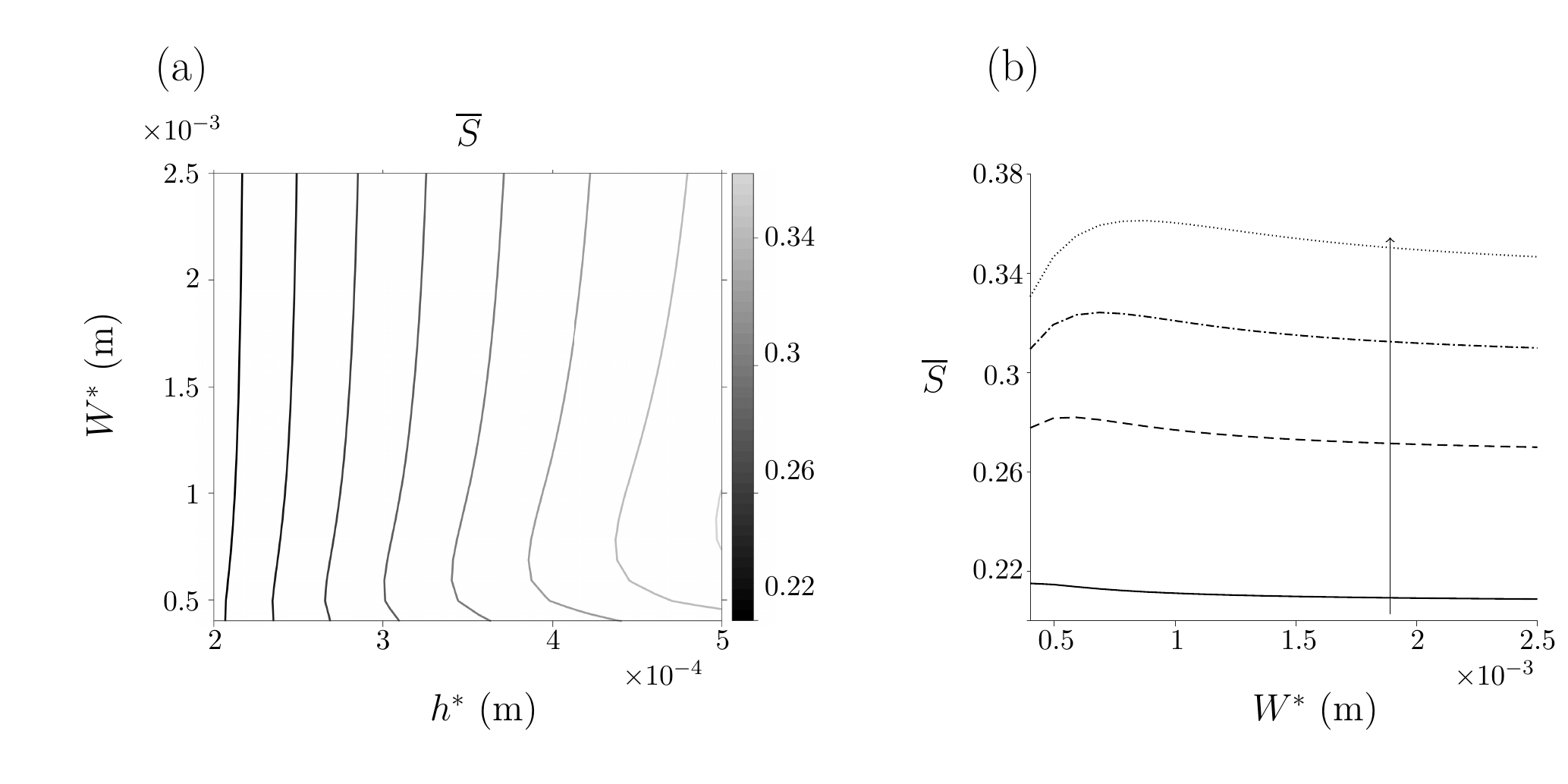}
\caption{The orientation parameter changing with channel dimensions for steady flow. (a) $\overline{S}$ for a range of channel width and depth, (b) $\overline{S}$ changing with increasing channel width $W^*$ for multiple small channel width values: $h^*=2\times10^{-4}~$m (solid line), $h^*=3\times10^{-4}~$m (dashed line), $h^*=4\times10^{-4}~$m (dot-dashed line) and $h^*=5\times10^{-4}~$m (dotted line) and the arrow denotes increasing channel width. Here $G^*=4.4\times 10^3~$Pa/m and $n_d^*=7.33\times 10^{17}~$phage/m$^3$.}
\label{figure7}
\end{figure}

\begin{figure}[!h]
\centering
\includegraphics[scale=0.63, trim=1.1cm 0.6cm 0cm 0.7cm,clip]{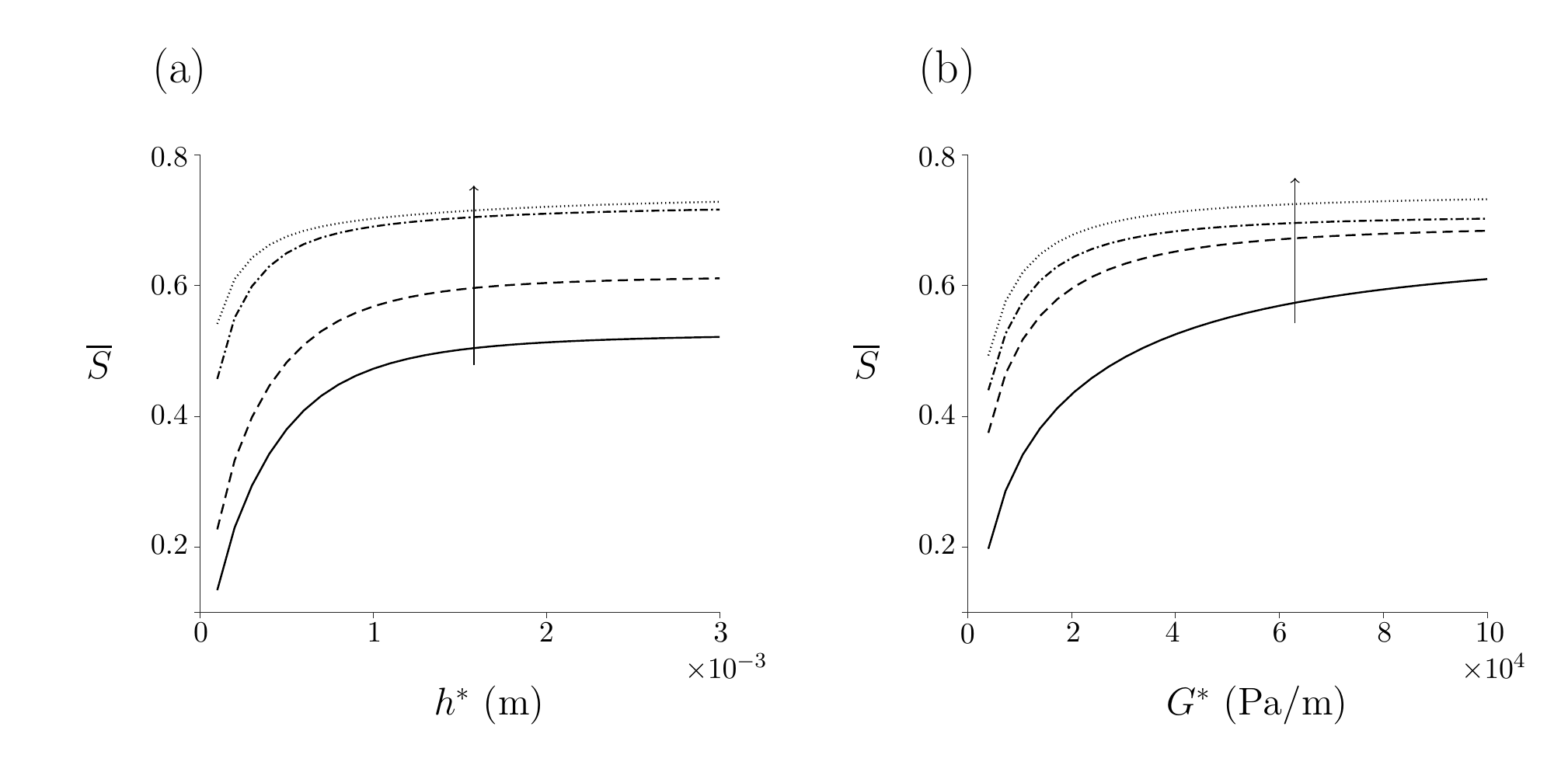}
\caption{Limit of $\overline{S}$ with increasing channel depth and pressure gradient for steady flow. (a) $\overline{S}$ with varying channel depth $h^*$ for a range of pressure gradients: $G^*=5\times 10^3~$Pa/m (solid line), $G^*=1\times10^4~$Pa/m (dashed line), $G^*=5\times10^4~$Pa/m (dot-dashed line) and $G^*=10\times 10^4~$Pa/m (dotted line). The arrow denotes increasing $G^*$. (b) $\overline{S}$ varying with pressure gradient $G^*$ for a range of channel depths: $h^*=2\times10^{-4}~$m (solid line), $h^*=6\times10^{-4}~$m (dashed line), $h^*=1\times10^{-3}~$m (dot-dashed line) and $h^*=4\times10^{-3}~m$ (dotted line). The arrow denotes increasing $h^*$.}
\label{figure8} 
\end{figure}

The spatially averaged orientation parameter is calculated for a range of channel width and depth in figure \ref{figure7}; the other parameters, including pressure gradient, remain fixed. The channel depth has a large impact on $\overline{S}$; for a deeper channel, the orientation parameter is larger. The channel width has limited effect, unless the two dimensions are comparable (figure \ref{figure7} (b)). For channel widths below $W^*=1\times10^{-3}~$m increasing the channel width increases the orientation parameter; this effect reverses as $W^*$ grows. Finally, the upper asymptote of the spatially averaged orientation parameter is investigated for large pressure gradient and channel depth (figure \ref{figure8}). In both cases, $\overline{S}$ quickly tends towards a constant value, $\overline{S}\approx0.74$.

\subsection{Oscillating pressure gradient}

\begin{figure}[!h]
\centering
\includegraphics[scale=0.7, trim=2.5cm 0.7cm 0cm 0.6cm, clip]{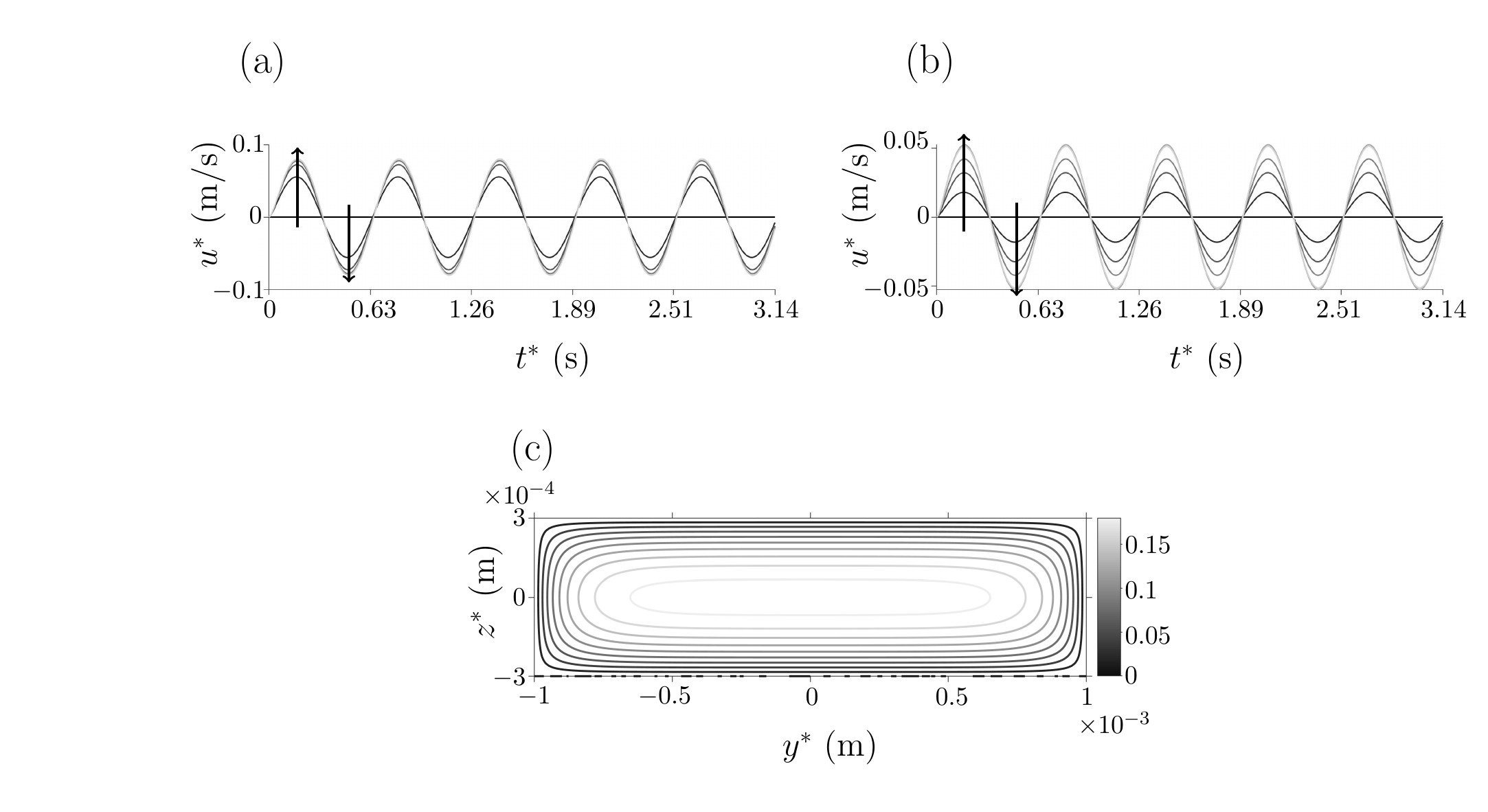}
\caption{Velocity field for oscillating flow and the standard parameter set. (a)  The velocity changing over time for multiple $y^*$ where $z^*=3\times10^{-4}~$m. The arrows denote increasing positive values of $y^*$. (b) The velocity changing over time for multiple points in $z^*$, where $y^*=2\times10^{-3}~$m. The arrows denote increasing positive values of $z^*$. (c) The velocity field, averaged over half a time period, through the cross-sectional face of the channel.}
\label{figure9}
\end{figure}

\begin{figure}[!h]
\centering
\includegraphics[scale=0.55, trim=1cm 0.3cm 0cm 0.7cm, clip]{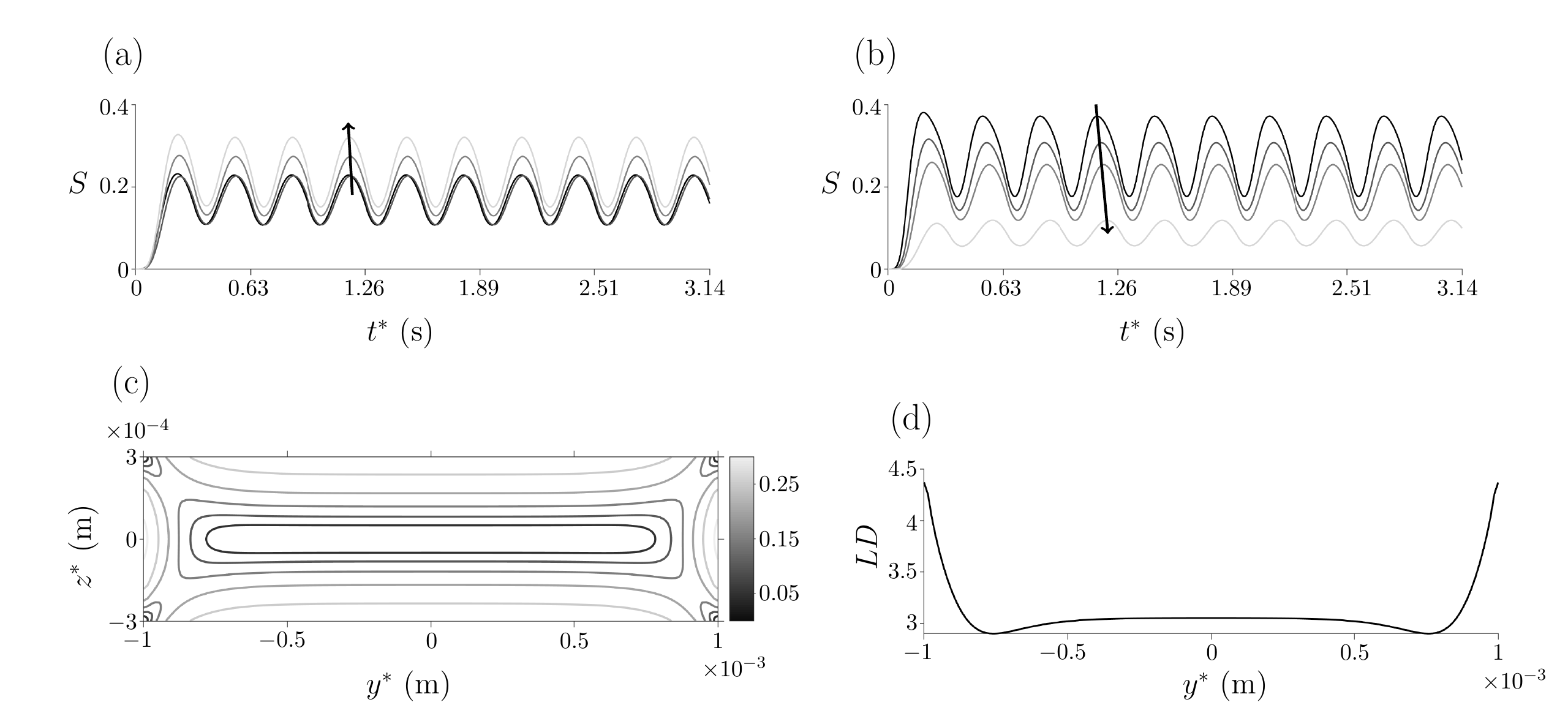}
\caption{The orientation parameter for oscillating flow and the standard parameter set. (a) $S$ evolving in time for different values of $z^*$ where $y^*=1.9\times10^{-3}~$m. The arrow denotes increasing $z^*$. (b) $S$ evolving in time for different values of $z^*$ where $y^*=3.5\times10^{-4}~$m. The arrow denotes increasing $z^*$. (c) $S$ through the cross-sectional face of the channel, averaged over half a time period. (d) $LD$ averaged over half a time period.}
\label{figure10}
\end{figure}

The velocity, orientation parameter and linear dichroism signal for oscillatory flow with the standard parameter set are shown in figures \ref{figure9} and \ref{figure10}. Figures \ref{figure9} (a) and (b) show how the velocity changes in time across the width and depth of the channel respectively; the velocity oscillates and is zero at the walls as expected. For figure \ref{figure9} (a), $z^*$ is chosen away from the walls to ensure non-zero velocity (and similarly for $y^*$ in figure \ref{figure9} (b)). A time averaged velocity profile is plotted in figure \ref{figure9} (c), this is qualitatively similar to the steady velocity profile in figure \ref{figure2} (a) but the magnitude of the velocity is reduced.

The orientation parameter is shown over time and for multiple values of $z^*$ in figures  \ref{figure10} (a), for $y^*=1.9\times10^{-3}~$m, and \ref{figure10} (b), for $y^*=3.5\times10^{-4}~$m. Towards the middle width (figure \ref{figure10} (a)), the orientation parameter increases with the channel depth and in the centre of the channel (figure \ref{figure10} (b)) the opposite occurs. Figure \ref{figure10} (c) displays the time averaged orientation parameter which agrees well with that produced in steady flow (figure \ref{figure2} (b)); this is also seen for the linear dichroism signal in figure \ref{figure10} (d). Note here that the peak values of the orientation parameter are similar for steady and oscillatory flow.

\renewcommand{\arraystretch}{1.1}
\begin{table} 
\begin{center}
\begin{tabular}{c c c c c c}
\hline
$G^*$ & $h^*$ & $\overline{S}$  & $\overline{S}$  \\
	  &		  & {\scriptsize ($\omega^*=10~$s$^{-1}$)} & {\scriptsize ($\omega^*=100~$s$^{-1}$)} \\
\hline
\multirow{2}{*}{$2\times10^3$} & $2\times10^{-4}$ & $0.130$ & $0.042$ \\[0.7ex]
						 	   & $4\times10^{-4}$ & $0.193$ & $0.031$ \\[0.7ex]
\multirow{2}{*}{$4\times10^3$} & $2\times10^{-4}$ & $0.241$ & $0.110$ \\[0.7ex]
						 	   & $4\times10^{-4}$ & $0.316$ & $0.081$ \\[0.7ex]
\multirow{2}{*}{$6\times10^3$} & $2\times10^{-4}$ & $0.309$ & $0.165$ \\[0.7ex]
						 	   & $4\times10^{-4}$ & $0.381$ & $0.122$ \\[0.7ex]
\multirow{2}{*}{$8\times10^3$} & $2\times10^{-4}$ & $0.353$ & $0.207$ \\[0.7ex]
						 	   & $4\times10^{-4}$ & $0.424$ & $0.154$ \\[0.7ex]
\end{tabular}
\end{center}
\caption{The spatially averaged orientation parameter for oscillating flow and a range of frequency of oscillations, pressure gradient and channel depth. Throughout these results, $n_d^*=1\times10^{18}~$phage/m$^3$ and $W^*=2\times10^{-3}~$m.}
\label{Table:LD_Gh}
\end{table}

\begin{figure}[!h]
\centering
\includegraphics[scale=0.7, trim=1.1cm 0.3cm 0cm 0.5cm, clip]{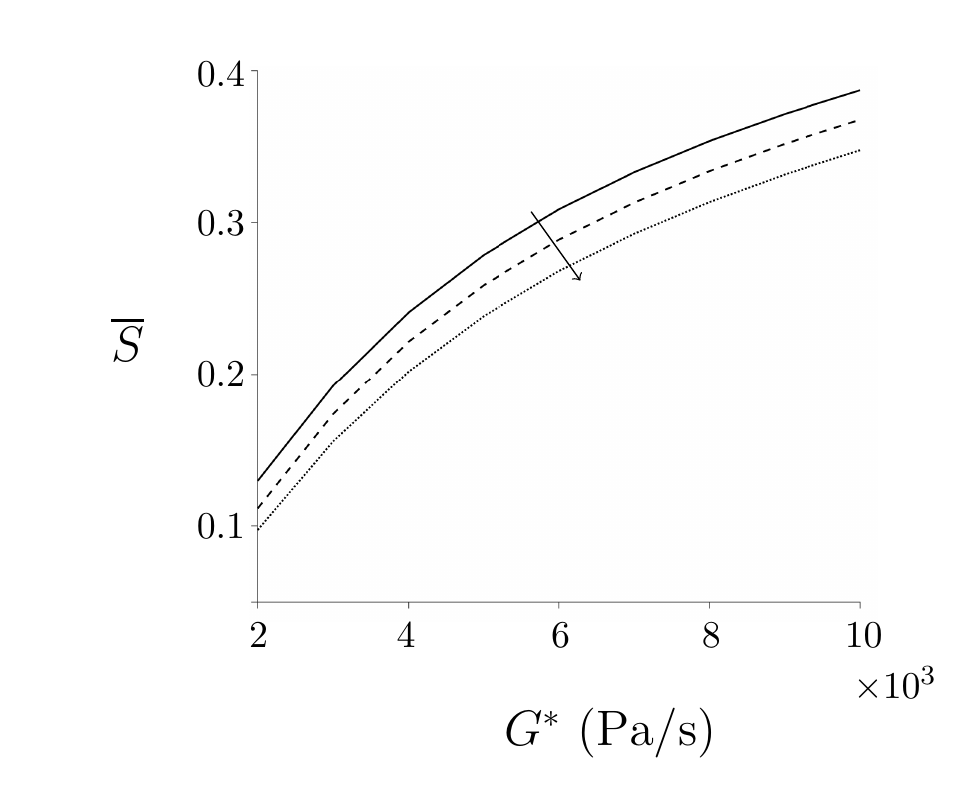}
\caption{Spatially-averaged orientation parameter, for oscillating flow, varying with increasing pressure gradient and for three values of $\omega^*$: $\omega^*=10~$s$^{-1}$ (solid line), $\omega^*=20~$s$^{-1}$ (dashed line) and  $\omega^*=30~$s$^{-1}$ (dotted line). Here $h^*=3\times 10^{-4}~$m, $W^*=1\times 10^{-3}~$m, $n_d^*=1\times10^{17}~$phage/m$^3$ and the arrow denotes increasing frequencies of oscillation.}
\label{figure11}
\end{figure}

Next, the spatially averaged orientation parameter $\overline{S}$ is compared for various pressure gradient and frequencies of oscillation (figure \ref{figure11}). As the pressure gradient $G^*$ increases, the orientation parameter increases; there is an indication that the impact of $G^*$ on $\overline{S}$ is reducing as $G^*$ becomes large. The impact of the frequency of oscillation $\omega^*$ is to decrease $\overline{S}$, a result seen throughout all figures. The relationship between the pressure gradient and frequencies of oscillation, for a choice of channel depth, is summarised in table \ref{Table:LD_Gh}. Changes in the number density have not been included for oscillating flow as the impact on the orientation parameter is small and reduces further for increasing frequencies of oscillations.

\begin{figure}[!h]
\centering
\includegraphics[scale=0.7, trim=1cm 0.4cm 0cm 0.6cm, clip]{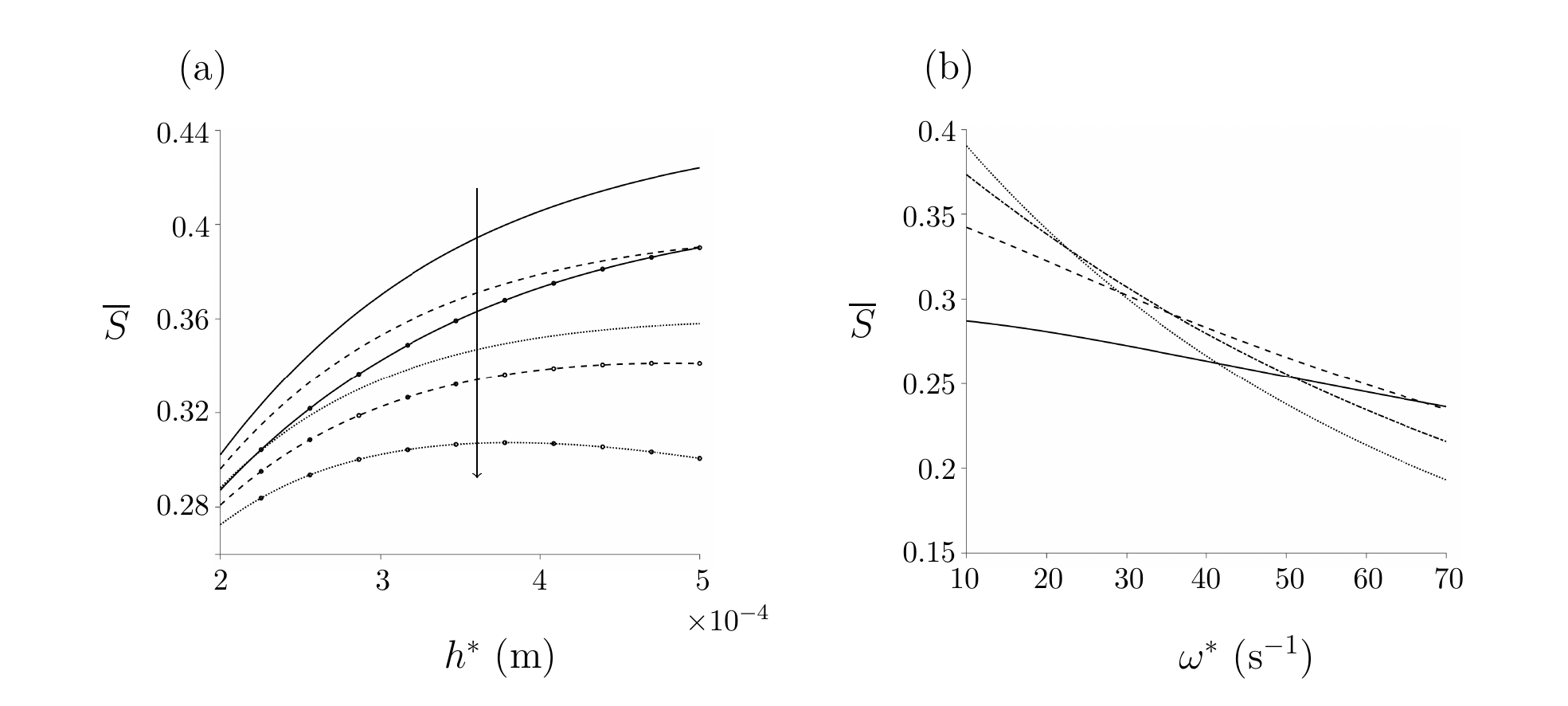}
\caption{Spatially averaged orientation parameter for oscillating flow, changing with channel dimensions and frequency of oscillation. (a) $\overline{S}$ varying with channel depth for multiple frequencies of oscillation: $\omega^*=10~$s$^{-1}$ (solid line), $\omega^*=20~$s$^{-1}$ (dashed line) and  $\omega^*=30~$s$^{-1}$ (dotted line). The lines without markers are for $W^*=5\times10^{-4}~$m and the lines with markers are for $W^*=2\times10^{-3}~$m. The arrow denotes increasing frequencies of oscillations. (b) $\overline{S}$ changing with frequency of oscillations for multiple values of $h^*$: $h^*= 2\times10^{-4}~$m (solid line), $h^*=3\times10^{-4}~$m (dashed line), $h^*=4\times10^{-4}~$m (dot-dashed line) and $h^*=5\times10^{-4}~$m (dotted line) and for $W^*=2\times10^{-3}~$m. For both figures, $G^*=8\times10^3~$Pa/m and $n_d^*=1\times10^{17}~$phage/m$^3$.}
\label{figure12}
\end{figure}

The orientation parameter is calculated for increasing channel depth $h^*$ and for a number of frequencies of oscillation and channel width (figure \ref{figure12}). Increasing $W^*$, shown by line markers in figure \ref{figure12} (a), decreases the orientation parameter in general. The orientation parameter has a non-monotonic response to $h^*$as the frequency of oscillations is increased; for small $\omega^*$, increasing $h^*$ increases the orientation parameter whereas for large $\omega^*$ this relationship begins to reverse. This trend is highlighted in figure \ref{figure12} (b), where $W^*=2\times10^{-3}~$m. The relationship between $h^*$, $W^*$, $\omega^*$ and the orientation parameter $\overline{S}$ has been summarised in table \ref{Table:LD_range}; when $\omega^*$ is large, a larger reduction in $\overline{S}$ is seen with increasing channel width. 

\begin{figure}[!h]
\centering
\includegraphics[scale=0.7, trim=1cm 0.3cm 0cm 0.5cm, clip]{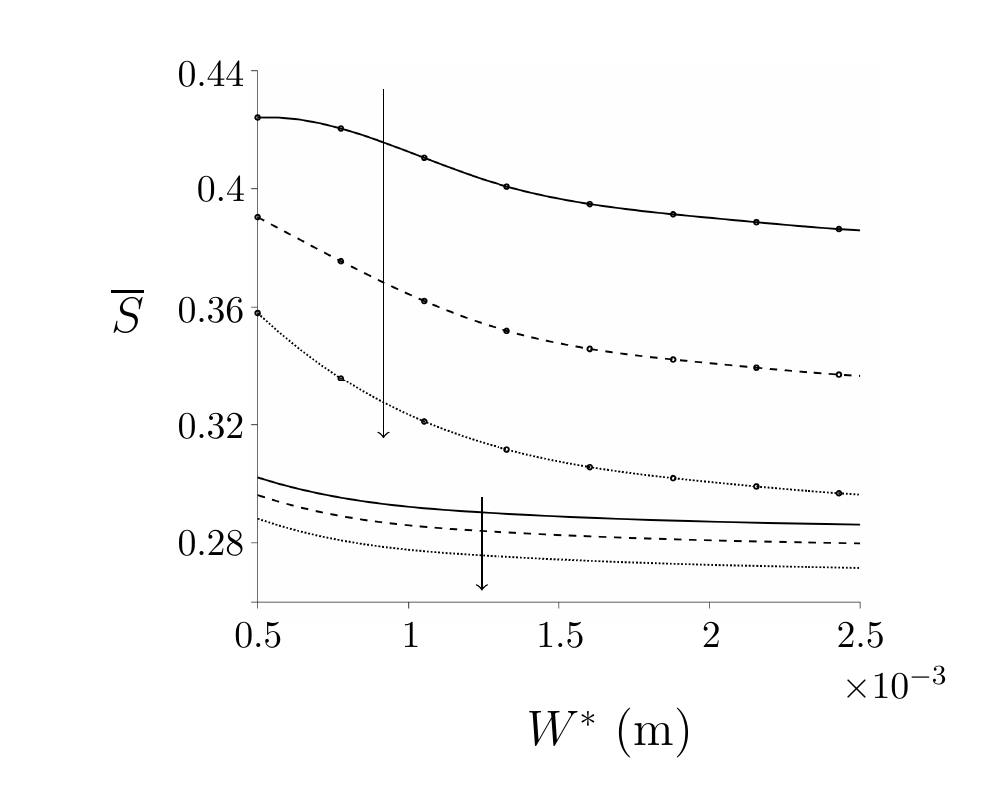}
\caption{Spatially averaged orientation parameter, for oscillating flow, changing with increasing channel width $W^*$. Three values of the frequency of oscillations are plotted: $\omega^*=10~$s$^{-1}$ (solid lines), $\omega^*=20~$s$^{-1}$ (dashed lines) and $\omega^*=30~$s$^{-1}$ (dotted lines). Two values of $h^*$ are also displayed: $h^*=2\times 10^{-4}~$m is shown by plain lines and $h^*=5\times10^{-4}~m$ is shown by the line markers. The arrows denote increasing frequencies of oscillation.}
\label{figure13}
\end{figure}

To further investigate these properties, the orientation parameter is plotted for increasing channel width for a range of channel depth and frequencies of oscillations (figure \ref{figure13}). As seen previously, the orientation parameter $\overline{S}$ increases with channel depth and decreases with increasing channel width. Increasing the frequency of oscillations reduces the orientation parameter in general; for larger channel depths, the decrease in the orientation parameter with $\omega^*$ is greater.

\renewcommand{\arraystretch}{1.2}
\begin{table} 
\begin{center}
\begin{tabular}{c c c c c}
\hline
$W^*$ & $h^*$ & $\overline{S}$ & $\overline{S}$ \\
	  &		  & {\scriptsize ($\omega^*=10~$s$^{-1}$)} & {\scriptsize ($\omega^*=100~$s$^{-1}$)} \\
\hline
\multirow{2}{*}{$0.5\times10^{-3}$} & $2\times10^{-4}$ & $0.302$ & $0.227$ \\[0.7ex]
							 	    & $4\times10^{-4}$ & $0.406$ & $0.217$ \\[0.7ex]
\multirow{2}{*}{$2\times10^{-3}$}   & $2\times10^{-4}$ & $0.287$ & $0.212$ \\[0.7ex]
							 	    & $4\times10^{-4}$ & $0.373$ & $0.173$ \\[0.7ex]
\end{tabular}
\end{center}
\caption{The spatially averaged orientation parameter for oscillating flow and a range of channel width, number density and frequencies of oscillations. Throughout these results, $G^*=8\times10^3~$Pa/m and $n_d^*=1\times10^{18}~$phage/m$^3$.}
\label{Table:LD_range}
\end{table}

\section{Discussion} \label{Sec:Disc}

The coupled pressure-driven flow and orientation dynamics of a dilute suspension of elongated particles has been modelled, to provide a mechanistic underpinning for flow linear dichroism spectroscopy. The model was applied specifically to a prototype hand-held device for waterborne pathogen detection, to explore the effect of changing the channel dimensions and pressure gradient, and to assess the feasibility of oscillatory flow. Mathematically, the problem involved coupling the Navier-Stokes equations, simplified via lubrication theory, to the Fokker-Planck equation for the shear-induced orientation dynamics. Whereas the flow problem involved two spatial independent variables and time, the orientation problem involved two angular independent variables and time, applied at each spatial location in the domain; the latter follows from the assumption that the particles are sufficiently small that they encounter a locally uniform shear rate. The flow and orientation problems were coupled via corrections to the fluid constitutive law involving moments of the orientation distribution, as described by established orientation Brownian suspension mechanics theory. 

The system was solved numerically via spherical harmonics for the orientation distribution and a finite difference method for the fluid dynamics, with iterative coupling; the oscillatory problem was temporally discretised via a second-order explicit method for orientation dynamics and an alternating direction implicit method for the fluid dynamics. The oscillatory flow is rather more computationally expensive due to temporal dependence; a parameter set of 175 tuples was explored for the oscillatory flow, with each tuple requiring 14--20 hours of walltime on 5 workstation cores. The dependence of particle orientation on channel width, depth, pressure gradient, particle number density and frequency of oscillation were reported and compared for both the steady and oscillatory systems.

To model linear dichroism, the analysis focused on a single orientation parameter, calculated as the average of the difference between the components of the particle alignment parallel to, and perpendicular to, the flow direction. It was noted that the earlier (homogeneous shear) analysis of McLachlan \cite{mclachlan2013calculations} applied an alignment formula based on uni-axial orientation for elongated particles embedded in a membrane. However because flow linear dichroism produces a bi-axial orientation distribution -- in other words there is no rotational symmetry about the flow direction in general -- this earlier application was inappropriate. The correct bi-axial formula was compared with the uni-axial formula; it was found that the qualitative behaviour was similar, however the uni-axial definition overestimates the degree of orientation. 

In both steady and oscillatory flows, increasing the pressure gradient increases the shear rate and hence the degree of alignment in the system. For steady flow, a plateau was observed close to a -- perhaps maximum achievable -- value of \(\overline{S}=0.75\), above $G^*=5\times10^4~$Pa/s and $h^*=2\times10^{-3}~$m, whereas for oscillating flow the mean orientation did not exceed \(0.43\) for the range of parameters analysed. This result does however suggest that oscillatory flow may be a viable method for producing practically-useful levels of alignment, which in turn may be valuable in the analysis of small sample volumes. Alignment decays monotonically with driving frequency, however an increase from \(10\)~rad/s to \(40\)~rad/s results in less than a 25\% reduction in alignment. Increasing channel height has a significantly greater impact than increasing channel width, although for oscillating flow there is evidence of a local maximum in alignment as channel height is increased. For a given limited sample volume there is a clear trade-off between channel volume and oscillatory frequency; as channel volume is increased, a higher oscillatory frequency is needed to keep the sample underneath the detection window. The data reported should enable the optimal configuration for a given volume to be calculated.

Signal production was also modelled by multiplying the alignment parameter by the number density of particles; due to the relatively small impact of particle density on flow, the relationship between signal and number density was therefore approximately linear. While increasing the number density may therefore appear to be a simple way to optimise signal, in practice there are trade-offs with scattering -- which was not modelled in the present case -- and cost of the reagent. Optimising the remaining design parameters is therefore important. 

Avenues for further work include experimental testing of the predictions of how signal should vary with channel dimensions and pressure gradient, and direct testing of oscillating flow linear dichroism spectroscopy. It will also be important to take account of scattering and hence produce a more accurate model of how number density affects signal production. Other avenues include incorporating flexibility of particles -- since the persistence length of the M13 bacteriophage considered here is approximately \(1.3\)~\(\mu\)m \cite{khalil2007} and the fibres are of length \(0.8\)~\(\mu\)m a rigid fibre approximation is reasonable, however future work could incorporate the flexibility of the fibres.

While the present manuscript focused on flows of passive particle suspensions in thin rectangular channels, future work may include investigating the behaviour of an active suspension, for example involving self-motile elongated cells such as spermatozoa in pressure-driven flow or complex behaviours such as trapping in shear flow \cite{bearon2015trapping}. Active suspensions display collective behaviour \cite{hwang2014bioconvection, koch2011collective, saintillan2008instabilities, subramanian2011stability} and superfluidity \cite{lopez2015turning}. The model considered here can be augmented by an extra stress term to account for self-motile behaviour \cite{pedley1990new},
\begin{equation} \label{extra_stress_active}
\bs{\sigma}_A^*=\alpha_1\int_s\left(\bs{pp}-\frac{\textbf{I}}{3}\right)\psi\dif \bs{p}.
\end{equation}

The dynamics of suspensions of elongated particles provides an enduring subject in fluid dynamics, with novel application areas in continuing to be discovered. The challenging task of solving the complex models associated with non-homogeneous shear and time-dependent forcing is now becoming within reach of multicore computing hardware. We hope that the present manuscript provides both useful information in the design and optimisation of the technological system considered, and an effective framework which can be applied and extended to other systems in biophysical spectroscopy and cell analysis.


\subsection*{Funding}
This work was supported by Biotechnology and Biological Sciences Research Council Industrial CASE Studentship (BB/L015587/1) and Engineering and Physical Sciences Research Council Healthcare Technologies Challenge Award (Rapid Sperm Capture EP/N021096/1).

\subsection*{Acknowledgements}
The authors acknowledge Linear Diagnostics Limited for supporting this research. We thank Professor Tim Dafforn, Professor Alison Rodger, Dr Craig Holloway and Dr Meurig Gallagher for valuable discussion and help throughout this project.

\bibliographystyle{plain} 

\bibliography{3Dchannelbib}

\begin{thebibliography}{10}

\bibitem{aubrey1991raman}
KL~Aubrey and GJ~Thomas~Jr.
\newblock Raman spectroscopy of filamentous bacteriophage {F}f (fd, {M}13, f1)
  incorporating specifically-deuterated alanine and tryptophan side chains.
  {A}ssignments and structural interpretation.
\newblock {\em Biophys J}, 60(6):1337--1349, 1991.

\bibitem{batchelor1970stress}
GK~Batchelor.
\newblock The stress system in a suspension of force-free particles.
\newblock {\em J Fluid Mech}, 41(3):545--570, 1970.

\bibitem{batchelor1971stress}
GK~Batchelor.
\newblock The stress generated in a non-dilute suspension of elongated
  particles by pure straining motion.
\newblock {\em J Fluid Mech}, 46(4):813--829, 1971.

\bibitem{bearon2015trapping}
RN~Bearon and AL~Hazel.
\newblock The trapping in high--shear regions of slender bacteria undergoing
  chemotaxis in a channel.
\newblock {\em J Fluid Mech}, 771, 2015.

\bibitem{bird1977dynamics}
RB~Bird, RC~Armstrong, O~Hassager, and CF~Curtiss.
\newblock {\em Dynamics of polymeric liquids}, volume~1.
\newblock Wiley New York, 1977.

\bibitem{bulheller2007circular}
BM~Bulheller, A~Rodger, and JD~Hirst.
\newblock Circular and linear dichroism of proteins.
\newblock {\em Phys Chem Chem Phys}, 9(17):2020--2035, 2007.

\bibitem{creppy2016}
A~Creppy, F~Plourabou{\'e}, O~Praud, X~Druart, S~Cazin, H~Yu, and P~Degond.
\newblock Symmetry-breaking phase transitions in highly concentrated semen.
\newblock {\em J Royal Soc Interface}, 13(123), 2016.

\bibitem{cupples2017viscous}
G~Cupples, RJ~Dyson, and DJ~Smith.
\newblock Viscous propulsion in active transversely isotropic media.
\newblock {\em J Fluid Mech}, 812:501--524, 2017.

\bibitem{cupples2018viscous}
G~Cupples, RJ~Dyson, and DJ~Smith.
\newblock On viscous propulsion in active transversely isotropic media.
\newblock {\em J Fluid Mech}, 855:408--420, 2018.

\bibitem{daviter2013circular}
T~Daviter, N~Chmel, and A~Rodger.
\newblock Circular and linear dichroism spectroscopy for the study of
  protein--ligand interactions.
\newblock In {\em Protein-Ligand Interactions}, pages 211--241. Springer, 2013.

\bibitem{doi1978dynamics}
M~Doi and SF~Edwards.
\newblock Dynamics of rod--like macromolecules in concentrated solution. {P}art
  1.
\newblock {\em J Chem Soc Farad Trans 2}, 74:560--570, 1978.

\bibitem{douglas1955numerical}
J~Douglas and DW~Peaceman.
\newblock Numerical solution of two-dimensional heat-flow problems.
\newblock {\em AIChE J}, 1(4):505--512, 1955.

\bibitem{dyson2015investigation}
RJ~Dyson, JEF Green, JP~Whiteley, and HM~Byrne.
\newblock An investigation of the influence of extracellular matrix anisotropy
  and cell{\textendash}matrix interactions on tissue architecture.
\newblock {\em J Math Biol}, 72(7):1775--1809, 2016.

\bibitem{dyson2010fibre}
RJ~Dyson and OE~Jensen.
\newblock A fibre-reinforced fluid model of anisotropic plant cell growth.
\newblock {\em J Fluid Mech}, 655:472--503, 2010.

\bibitem{ericksen1960ti}
JL~Ericksen.
\newblock Transversely isotropic fluids.
\newblock {\em Colloid Polym Sci}, 173(2):117--122, 1960.

\bibitem{ezhilan2015transport}
B~Ezhilan and D~Saintillan.
\newblock Transport of a dilute active suspension in
  pressure{\textendash}driven channel flow.
\newblock {\em J Fluid Mech}, 777:482--522, 2015.

\bibitem{garab2009linear}
G~Garab and H~van Amerongen.
\newblock Linear dichroism and circular dichroism in photosynthesis research.
\newblock {\em Photosyn Res}, 101(2-3):135--146, 2009.

\bibitem{glucksman1992three}
MJ~Glucksman, S~Bhattacharjee, and L~Makowski.
\newblock Three-dimensional structure of a cloning vector: {X}-ray diffraction
  studies of filamentous bacteriophage {M}13 at 7 {\aa} resolution.
\newblock {\em J Mol Biol}, 226(2):455--470, 1992.

\bibitem{green2008extensional}
JEF Green and A~Friedman.
\newblock The extensional flow of a thin sheet of incompressible, transversely
  isotropic fluid.
\newblock {\em Eur J Appl Math}, 19(03):225--257, 2008.

\bibitem{guazzelli2011physical}
E~Guazzelli and JF~Morris.
\newblock {\em A physical introduction to suspension dynamics}.
\newblock Cambridge University Press, 2011.

\bibitem{hill2005bioconvection}
NA~Hill and TJ~Pedley.
\newblock Bioconvection.
\newblock {\em Fluid Dyn Res}, 37(1):1--20, 2005.

\bibitem{hinch1973time}
EJ~Hinch and LG~Leal.
\newblock Time{\textendash}dependent shear flows of a suspension of particles
  with weak {B}rownian rotations.
\newblock {\em J Fluid Mech}, 57(4):753--767, 1973.

\bibitem{hinch1975constitutive}
EJ~Hinch and LG~Leal.
\newblock Constitutive equations in suspension mechanics. {P}art 1. {G}eneral
  formulation.
\newblock {\em J Fluid Mech}, 71(03):481--495, 1975.

\bibitem{hinch1976constitutive}
EJ~Hinch and LG~Leal.
\newblock Constitutive equations in suspension mechanics. {P}art 2.
  {A}pproximate forms for a suspension of rigid particles affected by
  {B}rownian rotations.
\newblock {\em J Fluid Mech}, 76(01):187--208, 1976.

\bibitem{holloway2018influences}
CR~Holloway, G~Cupples, DJ~Smith, JEF Green, RJ~Clarke, and RJ~Dyson.
\newblock Influences of transversely isotropic rheology and translational
  diffusion on the stability of active suspensions.
\newblock {\em Roy Soc Open Sci}, 5(8):180456, 2018.

\bibitem{holloway2015couette}
CR~Holloway, RJ~Dyson, and DJ~Smith.
\newblock Linear {T}aylor{\textendash}{C}ouette stability of a transversely
  isotropic fluid.
\newblock {\em Proc R Soc Lond A}, 471(2178), 2015.

\bibitem{holloway2018linear}
CR~Holloway, DJ~Smith, and RJ~Dyson.
\newblock Linear {R}ayleigh--{B}enard stability of a transversely isotropic
  fluid.
\newblock {\em Eur J Appl Math}, pages 1--23, 2018.

\bibitem{hwang2014bioconvection}
Y~Hwang and TJ~Pedley.
\newblock Bioconvection under uniform shear: linear stability analysis.
\newblock {\em J Fluid Mech}, 738:522--562, 2014.

\bibitem{jeffery1922motion}
GB~Jeffery.
\newblock The motion of ellipsoidal particles immersed in a viscous fluid.
\newblock {\em Proc R Soc Lond A}, 102(715):161--179, 1922.

\bibitem{kamal1989prediction}
MR~Kamal and AT~Mutel.
\newblock The prediction of flow and orientation behavior of short fiber
  reinforced melts in simple flow systems.
\newblock {\em Polym Composite}, 10(5):337--343, 1989.

\bibitem{khalil2007}
A.S. Khalil, J.M. Ferrer, R.R. Brau, S.T. Kottmann, C.J. Noren, M.J. Lang, and
  A.M. Belcher.
\newblock Single m13 bacteriophage tethering and stretching.
\newblock {\em Proc Natl Acad Sci}, 104(12):4892--4897, 2007.

\bibitem{kim2013microhydrodynamics}
S~Kim and SJ~Karrila.
\newblock {\em Microhydrodynamics: principles and selected applications}.
\newblock Courier Corporation, 2013.

\bibitem{koch2011collective}
DL~Koch and G~Subramanian.
\newblock Collective hydrodynamics of swimming microorganisms: {L}iving fluids.
\newblock {\em Annu Rev Fluid Mech}, 43:637--659, 2011.

\bibitem{leal1972rheology}
LG~Leal and EJ~Hinch.
\newblock The rheology of a suspension of nearly spherical particles subject to
  {B}rownian rotations.
\newblock {\em J Fluid Mech}, 55(04):745--765, 1972.

\bibitem{leslie1968some}
FM~Leslie.
\newblock Some constitutive equations for liquid crystals.
\newblock {\em Arc Rational Mech Anal}, 28(4):265--283, 1968.

\bibitem{lopez2015turning}
HM~L{\'o}pez, J~Gachelin, C~Douarche, H~Auradou, and E~Cl{\'e}ment.
\newblock Turning bacteria suspensions into superfluids.
\newblock {\em Phys Rev Lett}, 115(2):028301, 2015.

\bibitem{manhart2003rheology}
M~Manhart.
\newblock Rheology of suspensions of rigid{\textendash}rod like particles in
  turbulent channel flow.
\newblock {\em J Non-Newton Fluid}, 112(2):269--293, 2003.

\bibitem{marchioli2010orientation}
C~Marchioli, M~Fantoni, and A~Soldati.
\newblock Orientation, distribution, and deposition of elongated, inertial
  fibers in turbulent channel flow.
\newblock {\em Phys Fluids}, 22(3):033301, 2010.

\bibitem{marrington2005validation}
R~Marrington, TR~Dafforn, DJ~Halsall, JI~MacDonald, M~Hicks, and A~Rodger.
\newblock Validation of new microvolume {C}ouette flow linear dichroism cells.
\newblock {\em Analyst}, 130(12):1608--1616, 2005.

\bibitem{mclachlan2013calculations}
JRA McLachlan, DJ~Smith, NP~Chmel, and A~Rodger.
\newblock Calculations of flow-induced orientation distributions for analysis
  of linear dichroism spectroscopy.
\newblock {\em Soft Matter}, 9(20):4977--4984, 2013.

\bibitem{mortensen2008dynamics}
PH~Mortensen, HI~Andersson, JJJ Gillissen, and BJ~Boersma.
\newblock Dynamics of prolate ellipsoidal particles in a turbulent channel
  flow.
\newblock {\em Phys Fluids}, 20(9):093302, 2008.

\bibitem{niu2008bacteriophage}
Z~Niu, MA~Bruckman, B~Harp, CM~Mello, and Q~Wang.
\newblock Bacteriophage {M}13 as a scaffold for preparing conductive polymeric
  composite fibers.
\newblock {\em Nano Res}, 1(3):235--241, 2008.

\bibitem{pacheco2011detection}
R~Pacheco-G{\'o}mez, J~Kraemer, S~Stokoe, HJ~England, CW~Penn, E~Stanley,
  A~Rodger, J~Ward, MR~Hicks, and TR~Dafforn.
\newblock Detection of pathogenic bacteria using a homogeneous immunoassay
  based on shear alignment of virus particles and linear dichroism.
\newblock {\em Anal Chem}, 84(1):91--97, 2011.

\bibitem{pedley1990new}
TJ~Pedley and JO~Kessler.
\newblock A new continuum model for suspensions of gyrotactic micro-organisms.
\newblock {\em J Fluid Mech}, 212:155--182, 1990.

\bibitem{peterlin1939}
A.~Peterlin and H.A. Stuart.
\newblock Zur theorie der str{\"o}mungsdoppelbrechung von kolloiden und
  gro{\ss}en molek{\"u}len in l{\"o}sung.
\newblock {\em Z Phys}, 112(1-2):1--19, 1939.

\bibitem{pozrikidis2011fluid}
C~Pozrikidis.
\newblock {\em Introduction to theoretical and computational fluid dynamics}.
\newblock Oxford University Press, 2011.

\bibitem{norden1997circular}
A~Rodger and B~Nord{\'e}n.
\newblock {\em Circular dichroism and linear dichroism}.
\newblock Oxford University Press, USA, 1997.

\bibitem{saintillan2008instabilities}
D~Saintillan and MJ~Shelley.
\newblock Instabilities, pattern formation, and mixing in active suspensions.
\newblock {\em Phys Fluids}, 20(12):123304, 2008.

\bibitem{strand1987computation}
SR~Strand, S~Kim, and SJ~Karrila.
\newblock Computation of rheological properties of suspensions of rigid rods:
  stress growth after inception of steady shear flow.
\newblock {\em J Non-Newton Fluid}, 24(3):311--329, 1987.

\bibitem{subramanian2011stability}
G~Subramanian, DL~Koch, and SR~Fitzgibbon.
\newblock The stability of a homogeneous suspension of chemotactic bacteria.
\newblock {\em Phys Fluids}, 23(4):041901, 2011.

\end{thebibliography}


\end{document}